\documentclass[12pt,preprint]{aastex}

\newcommand{\myemail}{eduardo.gonzalez@uah.es}
\newcommand{\kms}{{\hbox {km\thinspace s$^{-1}$}}}
\newcommand{\Lsun}{{\hbox {L$_\odot$}}}
\newcommand{\Msun}{{\hbox {M$_\odot$}}}
\newcommand{\arcsdot}{\rlap{.}''}
\newcommand{\cmt}{{\hbox {cm$^{-3}$}}}
\newcommand{\cmd}{{\hbox {cm$^{-2}$}}}
\newcommand{\hdo}{{\hbox {H$_{2}$O}}}
\newcommand{\nht}{{\hbox {NH$_{3}$}}}
\newcommand{\nhd}{{\hbox {NH$_{2}$}}}
\newcommand{\nhdos}{{\hbox {$n({\rm H}_{2})$}}}
\newcommand{\tk}{{\hbox {$T_{k}$}}}

\shorttitle{The far infrared spectrum of Arp 220}
\shortauthors{Gonz\'alez-Alfonso et al.}

\begin{document}


\title{The far-infrared spectrum of Arp 220\footnote{Based on observations 
with the Infrared Space Observatory, an ESA project with instruments funded 
by ESA Member States (especially the principal investigator countries: 
France, Germany, Netherlands, and the United Kingdom) and with the participation 
of ISAS and NASA.}}


\author{Eduardo Gonz\'alez-Alfonso\altaffilmark{1,2}}
\affil{Universidad de Alcal\'a de Henares, Departamento de F\'{\i}sica, 
Campus Universitario, E-28871 Alcal\'a de Henares, Madrid, Spain}
\email{\myemail}

\author{Howard A. Smith}
\affil{Harvard-Smithsonian Center for Astrophysics,
    60 Garden Street, Cambridge, MA 02138}
\email{hsmith@cfa.harvard.edu}

\author{Jacqueline Fischer}
\affil{Naval Research Laboratory, Remote Sensing Division, Code 7213,
Washington, DC 20375}
\email{jackie.fischer@nrl.navy.mil}

\and

\author{Jos\'e Cernicharo}
\affil{CSIC, IEM, Dpto. Astrof\'{\i}sica Molecular e Infrarroja, Serrano 123, 
E-28006 Madrid, Spain}
\email{cerni@damir.iem.csic.es}


\altaffiltext{1}{Visiting Astronomer, Harvard-Smithsonian Center for Astrophysics,
    60 Garden Street, Cambridge, MA 02138.}
\altaffiltext{2}{CSIC, IEM, Dpto. Astrof\'{\i}sica Molecular e Infrarroja, 
Serrano 123, E-28006 Madrid, Spain.}


\begin{abstract}
ISO/LWS grating observations of the ultraluminous infrared galaxy Arp 220
shows absorption in molecular lines of OH, H$_2$O, CH, NH, and NH$_3$, as
well as in the [O I] 63 $\mu$m line and emission in the [C II] 158 $\mu$m line.
We have modeled the continuum and the emission/absorption of all observed
features by means of a non-local radiative transfer code. The continuum from 
25 to 1300 $\mu$m is modeled as a warm (106 K) nuclear region that is
optically thick in the far-infrared, attenuated by an extended region 
(size 2$''$) that is heated mainly through absorption of nuclear infrared 
radiation. The molecular absorption in the nuclear region is 
characterized by high excitation due to the high infrared radiation 
density. The OH column densities are high toward the nucleus 
($2-6\times10^{17}$ cm$^{-2}$) and the extended region 
($\sim2\times10^{17}$ cm$^{-2}$). The H$_2$O column density is also high 
toward the nucleus ($2-10\times10^{17}$ cm$^{-2}$) and lower in the 
extended region. The column densities in a halo that accounts for the absorption 
by the lowest lying levels are similar to what are found in the diffuse clouds 
toward the star forming regions in the Sgr B2 molecular cloud complex near 
the Galactic Center. Most notable are the high column densities found for 
NH and NH$_3$ toward the nucleus, with values of $\sim1.5\times10^{16}$ cm$^{-2}$ 
and $\sim3\times10^{16}$ cm$^{-2}$, respectively, whereas the NH$_2$ column
density is lower than $\sim2\times10^{15}$ cm$^{-2}$. A combination of
PDRs in the extended region and hot cores 
with enhanced \hdo\ photodissociation and a possible shock contribution in the 
nuclei may explain the relative column densities of OH and \hdo, whereas the 
nitrogen chemistry may be strongly affected by cosmic ray ionization. 
The [C II] 158 $\mu$m line is well reproduced by our models and its
``deficit'' relative to the CII/FIR ratio in normal and starburst galaxies is 
suggested to be mainly a consequence of the dominant non-PDR component of 
far-infrared radiation, although our models alone cannot rule out extinction
effects in the nuclei.
\end{abstract}




\section{Introduction}

With a redshift of $z=0.018$, Arp 220 (IC 4553/4) is the nearest and 
one of the best studied ultraluminous infrared galaxies (ULIRGs). The tails 
observed in the optical, together with the double highly-obscured and compact 
($0\arcsdot3$) nuclei observed in the near and mid-infrared, as well as in
the millimeter, strongly suggest that the enormous luminosity of Arp 220, 
$\sim10^{12}$ \Lsun, is the result of galactic merging. Nevertheless, 
the concrete physical process responsible is still a matter of debate: 
the proposed sources are hidden active nuclei and/or bursts of star
formation.

Molecular observations of Arp 220 provide unique clues to the physical and
chemical processes ocurring in the nuclei and their surroundings. In the
millimeter region, CO observations have been carried out with
increasingly high
angular resolution \citep*{rad91,sco91,oku94,sco97,saka99}. In particular
they have shown that, on the one hand, CO emission arises from a region
significantly more extended than the nuclei ($\sim3-4''$), and on the other hand
that the CO(2-1) to CO(1-0) intensity ratio is lower than 1, thus suggesting
that CO mainly traces low density regions ($<10^3$ \cmt).
Observations of molecules with high dipole moment, like CS and HCN, have
revealed that the fraction of molecular gas contained in dense clouds
(\nhdos$\ge10^4$ \cmt) is much
larger than in normal galaxies, yielding $\sim10^{10}$ \Msun\ of dense
gas \citep*{sol90,sol92}. \citet{rad91b} found that HCN(1-0) and HCO$^+$(1-0) 
peak strongly toward the nuclei, but also show low-level extended emission.
More recently, \citet{aalto02} have detected emission from the high 
density tracers HNC and CN, and the relatively low HCN/CN and HCN/HNC 
intensity ratios were attributed to widespread PDR chemistry.

The launch of the Infrared Space Observatory (ISO) opened a new window
for the study of the physical and chemical properties of ultraluminous
infrared galaxies. Despite the lack of angular and spectral resolution,
the observations of Arp 220's far-infrared spectrum from 40 to 200
$\mu$m \citep{fis97,fis99} and of a number of individual lines in the SWS range 
\citep{stu96} provided new insights in our understanding of the ionic, atomic and
molecular content of the galaxy. These wavelength regions are of great
interest, because the bulk of the enormous luminosity is emitted in the
far-infrared, and also because they contain lines of important
molecular, ionic, and atomic
species. \citet{skin97} reported the detection of the 35 $\mu$m OH line
in Arp 220. \citet{fis97,fis99} found that the far-infrared molecular 
absorption lines of OH, \hdo, CH, and \nht\ are significantly stronger in
Arp 220 than in less luminous infrared-bright galaxies while the fine structure 
lines from ionic species are, to the contrary, extremely weak.
\citet{luh98,luh03} found that, like in the ULIRG Arp 220,
the [C II] to the far-infrared luminosity ratio 
is typically nearly an order of magnitude weaker than in lower
luminosity infrared-bright galaxies.  \citet{stu96} 
reported the detection of two ortho-H$_2$ pure rotational lines, indicating that 
high masses of gas are subject to PDR conditions and/or shock activity.
Although the few other ULIRGs for which either full LWS spectra or partial 
scans in the vicinity of the OH lines were obtained also show far-infrared 
molecular absorption \citep{keg99,har99}, the rich molecular LWS spectrum of 
Arp 220 is unique for its high signal-to-noise ratio.

The physical and chemical processes that account for the rich molecular
far-infrared spectrum of Arp 220 can be better understood if quantitative values 
of the column densities of the above species, as well as their excitation
conditions, are estimated. The presence of OH and, to some extent, of 
\hdo, may be indicative of PDR and/or diffuse interstellar cloud chemistry, 
and their column densities potentially give an estimate of the UV field in 
the source. On the other hand, large amounts of \hdo\ are produced in non-dissociative 
shocks \citep[e.g.][]{cerni1999}, where the OH abundance is also enhanced 
\citep{wat85}. 
The OH abundance is expected to be generally higher than that of \hdo\ in fast 
dissociative shocks \citep{neu89}. \hdo\ ice in grain mantles may also efficiently 
return to the gas phase through sublimation of mantles in ``hot core'' regions. But 
whatever the characteristics of the regions producing the observed molecular 
features, the lines under study lie at wavelengths where the enormous 
infrared continuum flux approaches its maximum, and the molecular excitation 
of high-dipole species should be strongly
affected by absorption of continuum radiation. Hence any reliable estimation of 
molecular column densities require accurate models for the dust emission. 
Unfortunately ISO's lack of angular resolution forces us to rely on 
plausibility arguments in our assumptions about the regions where the different 
lines are formed; some of these are based on general requirements of 
excitation, and others on conclusions from observations of galactic sources. 
The main goal of this work is thus to shed light on the physical and chemical 
processes in Arp 220, based on detailed model fits of its continuum and  
far-infrared molecular/atomic line absorption and emission spectrum. We adopt a 
distance to Arp 220 of 72 Mpc \citep[projected linear scale of 350 
pc/arcsec,][]{graham90}. In section \ref{sec:obs} we present the
ISO/LWS observations; in section \ref{sec:generalresults} we discuss the line
identifications; section \ref{sec:cont} is a discussion of the models for 
the continuum emission; section \ref{sec:modelmol} presents the models for the
molecular and atomic species; in section \ref{sec:discussion} we discuss the
implications of the radiative transfer models, and section \ref{sec:summary} 
summarizes our main results.

\section{Observations} \label{sec:obs}

The full 43-197 $\mu$m spectrum of Arp 220, obtained with the LWS
spectrometer \citep{clegg96} on board ISO \citep{kes96} (TDT\footnote{Target
Dedicated Time number, a unique identifier for each observation} number 27800202), 
was presented by \citet{fis97,fis99}. The grating spectral resolution is $\approx$0.3
$\mu$m in the 43--93 $\mu$m interval (detectors SW1--SW5), and of $\approx$0.6
in the 80--197 $\mu$m interval (detectors LW1--LW5), corresponding to
$\Delta v\ge10^3$ \kms. The lines are thus not resolved in velocity. The
beam size of $\approx80''$ ensures that all the continuum and line
emission/absorption from Arp 220 \citep[CO size $<4''$,][hereafter SYB97]{sco97}
lie within the ISO aperture.

The data we present here (coadded TDT numbers 27800202, 64000916, and
64000801) were reduced using version 9.1 of the Off
Line Processing (OLP) Pipeline system which we found to produce a higher 
signal-to-noise spectrum than OLP 10.1. However, we adopted the continuum 
correction given by OLP version 10.1, which typically gives absolute responsivity 
corrections with uncertainty factors $\sim3$ times lower than are produced by 
version 9 (Tim Grundy, private communication). 
We performed subsequent data processing, including further de-glitching, 
co-addition, and scaling, using the ISO Spectral Analysis Package 
\citep[ISAP;][]{sturm98}. In order to obtain a smooth spectrum 
throughout the whole LWS range, the flux densities given by each detector
were corrected by multiplicative scale factors. Corrections were lower 
than 10\% except for detector SW1 (43--64 $\mu$m), for which the
correction was of 15\%. Thus we attribute a conservative uncertainty
of 15\% to the overall continuum level.

\begin{figure*}
\epsscale{1}
\plotone{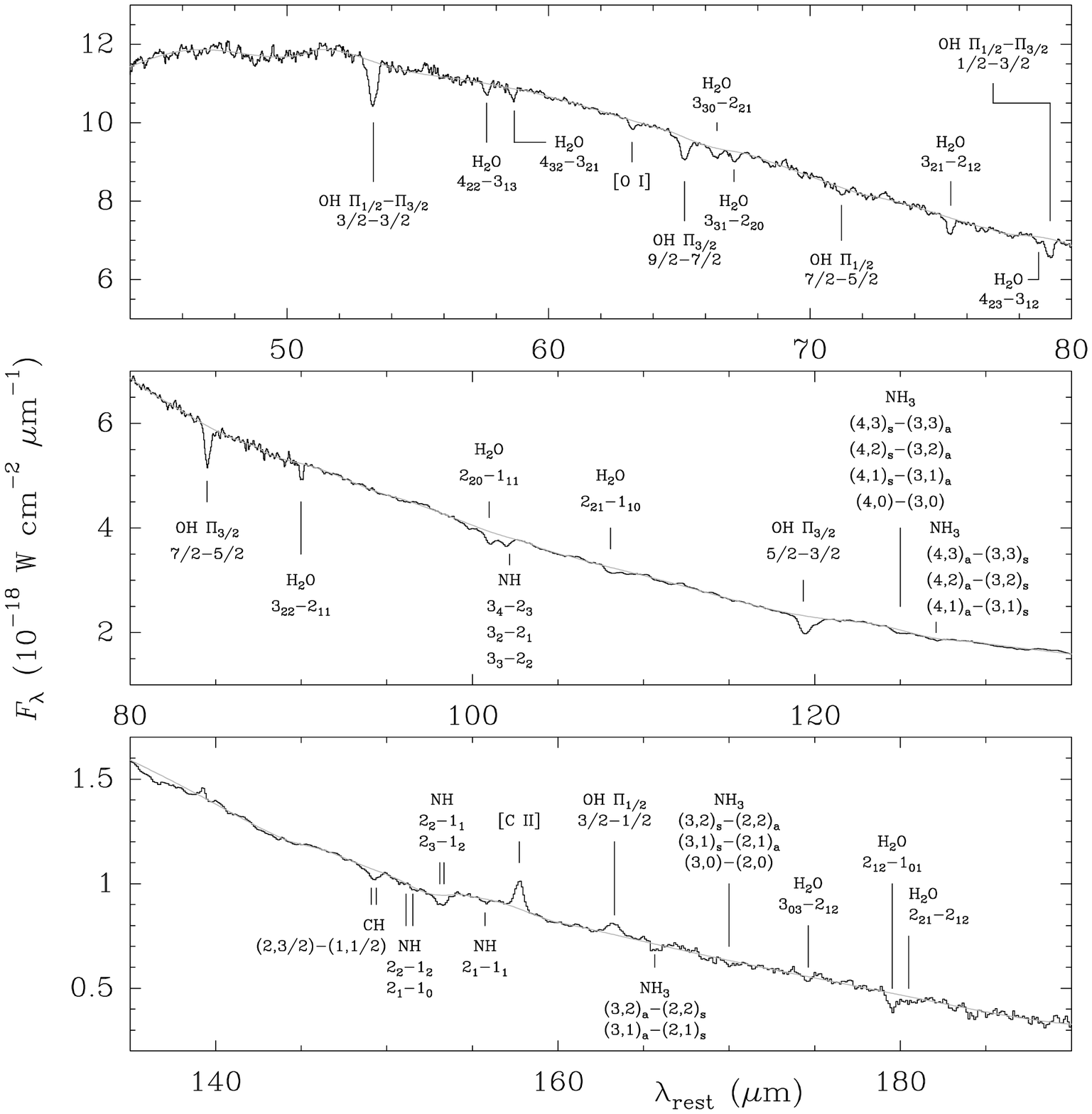}
\caption{ISO/LWS spectrum of Arp 220, where the most prominent line features
are identified (see text). The grey line shows the adopted baseline
(continuum level). Wavelengths in this and next figures are rest
wavelengths.
\label{fig:arp220}}
\end{figure*}

The LWS spectrum of Arp 220 is presented in Fig.~\ref{fig:arp220} together
with identifications of the most prominent lines.  Owing to transient effects, the 
fluxes of weak lines as observed in the forward and reverse scans were found to 
differ significantly in some wavelength ranges. In these cases, 
if a line appeared close to the upper or lower end of a detector, 
the reverse or forward scan was selected for that line, respectively, as has 
been found to minimize the transient effects (Tim Grundy, private communication). 
Nevertheless, the average of both scans was used throughout most of the
spectrum. In wavelength regions where two detectors' responses overlap, line fluxes 
were generally found to be consistent. The only exception was the \hdo\ 
$3_{22}-2_{11}$ line at 
90 $\mu$m, which showed in LW1 a flux 60\% weaker than in SW5. We adopted
here the SW5 spectrum, but the flux of the above \hdo\ line should be considered 
highly uncertain. The subtraction of a baseline (see Fig. \ref{fig:arp220}) added 
additional uncertainty to the line fluxes, particularly in cases of broad features 
presumably composed of several lines. With the exception of the \hdo\ $3_{22}-2_{11}$ 
line, we estimate a line flux uncertainty generally lower than 35\%.

We note for reference, that for the LWS spectral resolution $\Delta v=1500$ 
\kms, a line that is optically thick at line center with a line width of 300 
\kms\ and negligible line or continuum re-emission, would show aborption at 
line center of $\sim20$\% in Fig.~\ref{fig:arp220}.

\section{General results} \label{sec:generalresults}

\begin{figure*}
\epsscale{1.0}
\plotone{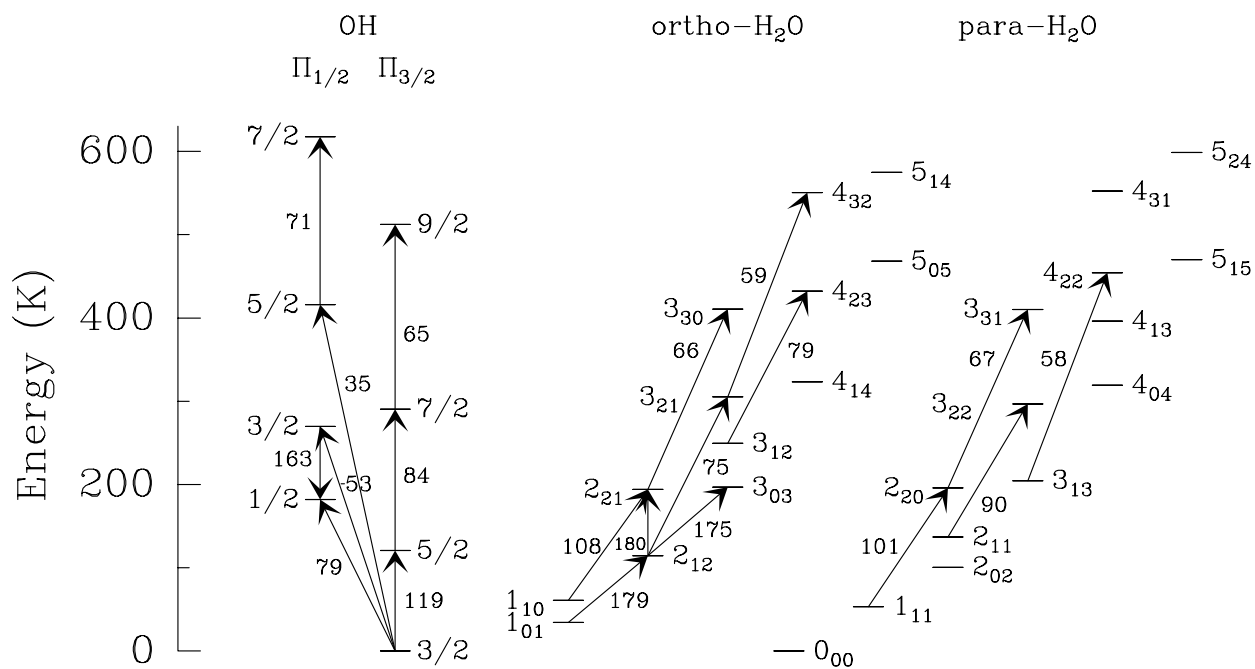}
\caption{Energy level diagrams of OH and \hdo\ (ortho and para). Rotational
levels with energies above the ground state up to 620 K are shown; the lines
detected in Arp 220 are indicated with arrows and their wavelengths are
in $\mu$m. OH $\Lambda$-doubling is ignored because the $\Lambda$-doublets 
are not resolved with the ISO grating resolution.}
\label{fig:levels}
\end{figure*}

{\it OH and \hdo:} The FIR spectrum of Arp 220 is dominated 
by unresolved OH doublets (hereafter simply referred to as lines)
and \hdo\ lines in absorption, with the exception of the OH
$\Pi_{1/2}\,3/2-1/2$ emission line at 163.3 $\mu$m. Figure
\ref{fig:levels} shows the level diagram of OH, ortho-\hdo\ and para-\hdo,
and indicates the lines detected in Arp 220. Lines with very different 
excitation requirements are observed throughout the spectrum. The OH lines
$\Pi_{3/2}\, J=9/2-7/2$ at 65 $\mu$m and $\Pi_{1/2} J=7/2-5/2$ at 71 $\mu$m
have lower levels at 290 K and 415 K above the ground state, respectively,
whereas the \hdo\ lines $4_{32}-3_{21}$ (59 $\mu$m) and $4_{22}-3_{13}$
(58 $\mu$m) have lower levels at 305 K and 205 K.
Strong absorption is also observed in the OH ground state lines at 53,
79, and 119 $\mu$m, as well as in the \hdo\ lowest-lying line at 179 $\mu$m.
This wide range of excitation suggests that several regions 
with different physical conditions are contributing to the observed features 
\citep{fis99}, and one of the goals of this work is to provide a
reasonable estimate of the nature of these regions and their relative
contributions to the spectrum. On the other hand, several of the lines have
complex shapes, with evidence of shoulders suggestive of weaker secondary lines.
In particular, the
ground state OH 119 $\mu$m line shows a redshifted ``shoulder'', which is
detected in both the forward and reverse scans, although with somewhat
different strengths. It could be attributed to the $^{18}$OH $\Pi_{3/2}\,5/2-3/2$
line at 120.1 $\mu$m, although contamination by other species such as 
CH$^+$ cannot be ruled out. 
Also, the redshifted ``wing'' of the \hdo\ $2_{12}-1_{01}$ line at 179 $\mu$m,
attributed in Fig. \ref{fig:arp220} to the \hdo\ $2_{21}-2_{12}$ line, could
also be contaminated by H$_2^{18}$O, CH (see Fig. \ref{fig:arp220model}),
and H$_3$O$^+$.

\begin{figure}
\epsscale{0.6}
\plotone{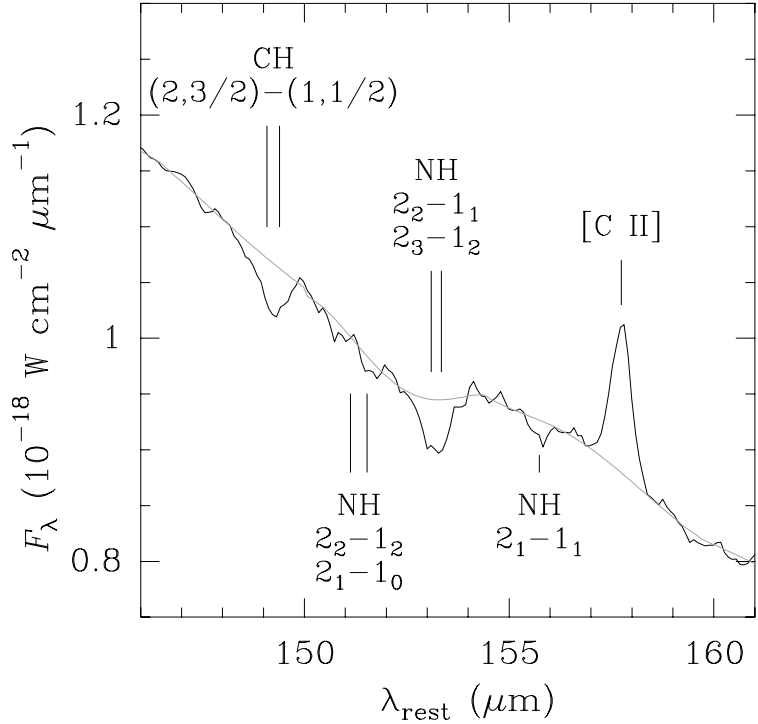}
\caption{Line assignments in the vecinity of the 153.2 $\mu$m feature. The
grey line shows the adopted continuum level.
\label{fig:nhlines}}
\end{figure}

{\it CH and NH:} The spectrum contains lines from other molecular species: 
like CH at 149 $\mu$m, NH$_3$ at 125, 166 and 170 $\mu$m and, very interestingly, 
strong absorptions at 102 and 153.2 $\mu$m that have been identified as NH in Fig 
\ref{fig:arp220}. Evidence for the latter identifications is strengthened
because of the presence of weak line-like features at 155.74 $\mu$m, and
marginally at 151.53 $\mu$m, which would correspond to the NH $2_1-1_1$
and $2_1-1_0$ lines, respectively (see also Fig. \ref{fig:nhlines}).
Conceivably, the line absorptions at 102 and 153.2 $\mu$m could be 
severely contaminated by other species, like C$_3$, H$_2^{18}$O, 
NH$_3$, and even OH$^+$. 
C$_3$ has a strong transition at 153.3 $\mu$m, but its contribution is expected to 
be minimal due to the lack of detection of other adjacent C$_3$ lines (in 
particular at 154.86 $\mu$m). The absorption at 102 $\mu$m may be contaminated 
by the H$_2^{18}$O $2_{20}-1_{11}$ line (just at 102.0 $\mu$m), but 
since the H$_2^{18}$O should not be as strong as the corresponding adjacent line 
of the main isotope H$_2^{16}$O, we regard this identification as also
unlikely. Some contribution of NH$_3$ lines to the 102 $\mu$m feature 
might be expected, but they are somewhat shifted in wavelength (they lie 
between 101.5 and 101.7 $\mu$m). Both features could be contaminated to some 
extent by OH$^+$, with strong lines at 101.70 and 101.92 $\mu$m ($N_J=3_3-2_2$ 
and $3_4-2_3$) and at 153.0 $\mu$m ($2_3-1_2$), but the strong
absorption at 153.2 $\mu$m absorption could never be
explained by OH$^+$ alone.  
Therefore, despite the possible contamination from other molecules, NH is 
probably responsible for most, if not all, of the observed 153.2 
$\mu$m absorption (see also Fig. \ref{fig:arp220model}). Thus, although 
the definitive assignment to NH should await confirmation with higher 
spectral resolution observations of the lowest-lying NH transitions at $\sim10^3$ 
GHz, we conclude that ISO observations strongly support its detection in Arp 220, 
and advance a model of the observed absorption (section \ref{sec:modelmol}) that can
be useful to direct future observational and theoretical studies. If confirmed, 
this detection is the first 
extragalactic detection of NH, which has been previously detected only in 
galactic diffuse clouds through electronic transitions \citep{meyer91,craw97} and, 
interestingly, toward Sgr B2 in the Galactic Center via the same 
transitions detected in Arp 220 \citep*[hereafter GRC04]{cerni2000,goi04}.

\begin{figure*}
\epsscale{1.0}
\plottwo{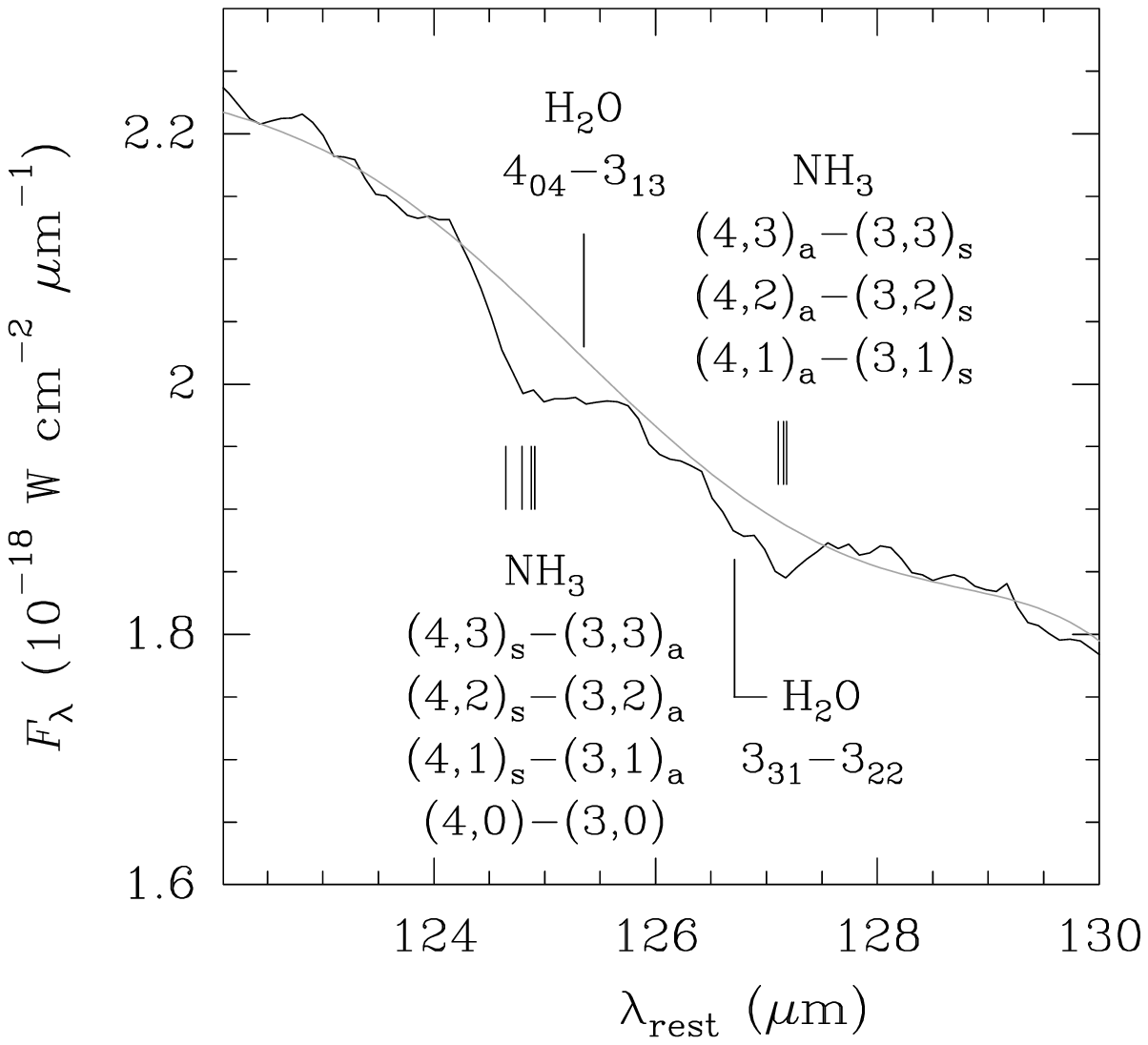}{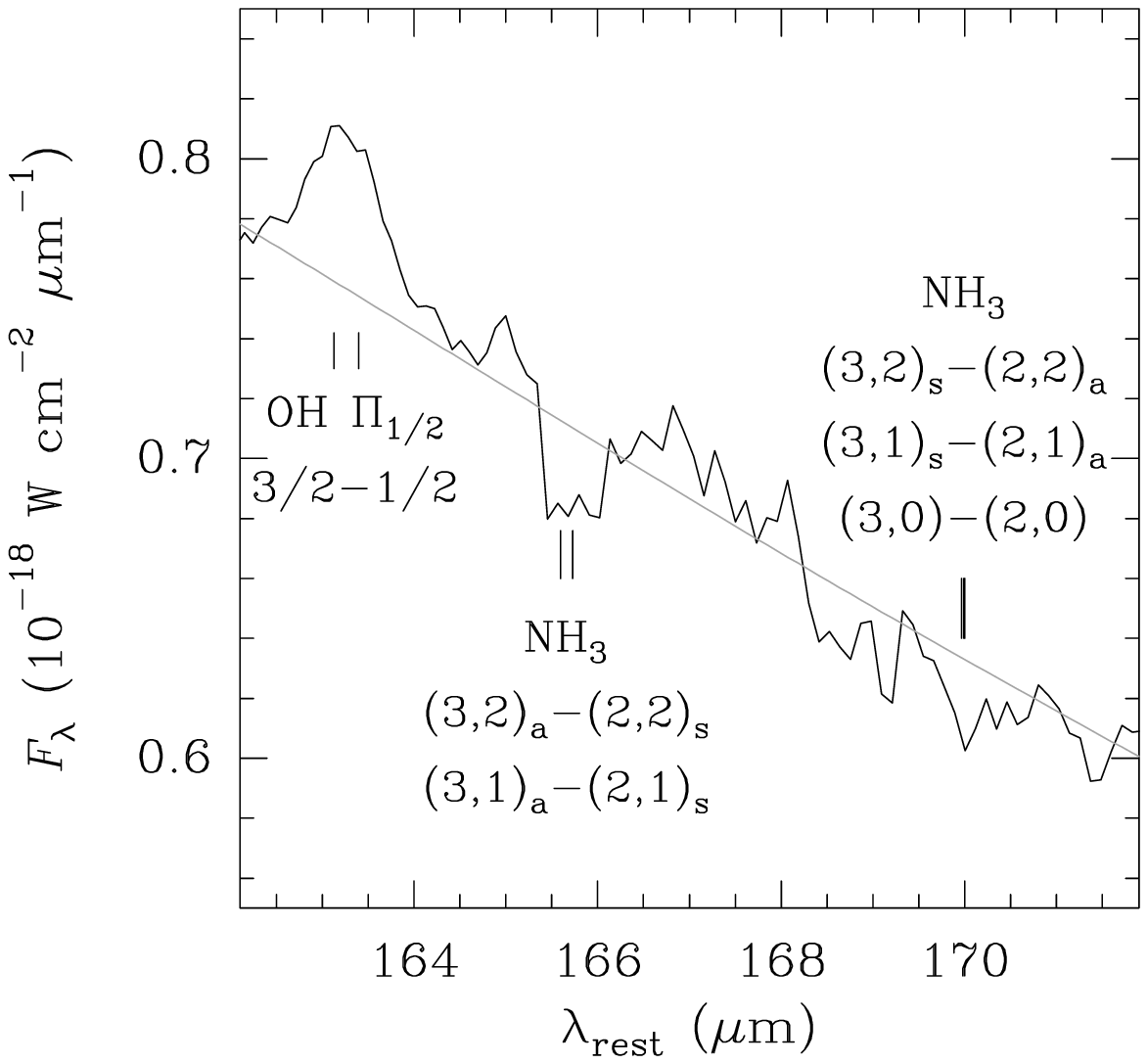}
\caption{NH$_3$ lines around 125, 127, 166, and 170 $\mu$m. The
grey line shows the adopted continuum level. The 125 and 127 $\mu$m features
could be partially contaminated by the labelled \hdo\ lines.
\label{fig:nh3lines}}
\end{figure*}

{\it NH$_3$:} The spectrum of Arp 220 around 125 and 170 $\mu$m, shown in
Figure \ref{fig:nh3lines}, strongly supports the identification of NH$_3$. 
The shape of the 165.7 $\mu$m feature indicates transient effects, but the line is
detected in both the forward and reverse scans. Some \hdo\ lines may
contribute to the observed 125 and 127 $\mu$m features, but they are
shifted in wavelength relative to the strongest absorption. This is the first
extragalactic detection of infrared NH$_3$ lines. With the detection of NH 
and NH$_3$ we might expect to detect NH$_2$, but there is no evidence for its 
strongest expected lines at 159.5, 117.8 and 104.9 $\mu$m. 

{\it CO?:} An apparent emission line, detected in both the forward and reverse
scans, is present at 173.7 $\mu$m in Fig. \ref{fig:arp220}. It coincides rather
 well with the expected position of the CO $J=15-14$ line at 173.63 $\mu$m. This 
identification cannot be confirmed by the detection of other expected CO lines because 
the CO $J=14-13$ line at 186.0 $\mu$m lies at the noisy edge of the LW5 detector and 
the CO $J=16-15$ line at 162.8 $\mu$m is blended with the OH 163.3 
$\mu$m line. The higher $J$ lines are expected to be too weak to be 
detectable, given the observed strength of the 173.7 $\mu$m feature.

{\it [O I] and [C II]:} The ISO spectrum also shows the [O I] 63.2 $\mu$m 
line in absorption and the [C II] 157.7 $\mu$m line in emission. The [O I] 145.5 
$\mu$m line is not detected. These lines, as observed in Arp 220 and other
ULIRGs, have been discussed elsewhere \citep{fis97,fis99,luh98,luh03}. In 
section \ref{sec:modelmol} we present a simple model of the Arp 220 spectrum
that may shed some light on the peculiar behavior of these lines in Arp 220.

In summary, the far-infrared spectrum of Arp 220 shows molecular lines of OH,
\hdo, CH, NH, and NH$_3$. The atomic lines of [O I] at 63 $\mu$m and [C II]
at 158 $\mu$m are also detected. Lines of other species, like CO, H$_3$O$^+$, 
CH$^+$, and OH$^+$, could also contaminate the observed features, but 
our limited spectral resolution prevents the possibility of unambiguous
detection. Only the [C II] 158 $\mu$m and the OH $\Pi_{1/2}\,3/2-1/2$ 163
$\mu$m lines are clearly observed in emission. Lines from ions that would
trace H II regions and/or an AGN are absent.

\section{Models for the continuum} \label{sec:cont}

Figure \ref{fig:arp220} shows that the continuum peaks around $\sim$ 40 -- 50
$\mu$m. The bulk of the continuum from Arp 220 is emitted by heated dust grains.
At 1.3 millimeter wavelengths, \citet{saka99} showed that the continuum arises
almost exclusively from the nuclei, with an equivalent size of $\sim0\arcsdot4$.
The non-thermal contribution at 1.3 mm is expected to be $<15$\% 
\citep[cf. Fig.~6 of][]{anant00}.
On the other hand, \citet[hereafter S99]{soifer99} have shown that the
two nuclei also account for essentially all the continuum at 25 $\mu$m.  
Combining both observations, S99 proposed two alternative scenarios to explain 
the continuum emission of Arp 220 from far-infrared to millimeter 
wavelengths. Our models of the continuum emission are entirely based on
these scenarios, which we have examined and refined quantitatively 
on the basis of our ISO 45-200 $\mu$m spectrum.

{\it Model S$_1$:} In the first scenario (hereafter S$_1$), it is assumed that the 
emission from the nuclei is not significantly attenuated at 25 $\mu$m by 
foreground material. We have 
simulated the emission from the nuclei as arising from a single nucleus with effective 
size of $0\arcsdot41$. With an effective dust temperature of 85 K, and optically thick 
emission in the submillimeter, the requirement that the fluxes at 25 and 1300 
$\mu$m arise from the nuclei is fulfilled. However, the emission at 60-100 $\mu$m, 
as well as the total luminosity from the galaxy, are then underestimated and a more 
extended region (hereafter ER) must be invoked to account for the remaining flux. 
We identify this surrounding environment with the extended emission observed in 
CO, HCN, and HCO$^+$. In this first scenario, then, most of the Arp 220 
luminosity is produced in the ER, which has been modeled as a thin disk by 
SYB97, and as a warped disk by \citet{eck01}. Significantly, if this model is 
correct, a spatially extended starburst, responsible for the bulk of the far-infrared
luminosity is inferred.

\begin{figure}
\epsscale{0.8}
\plotone{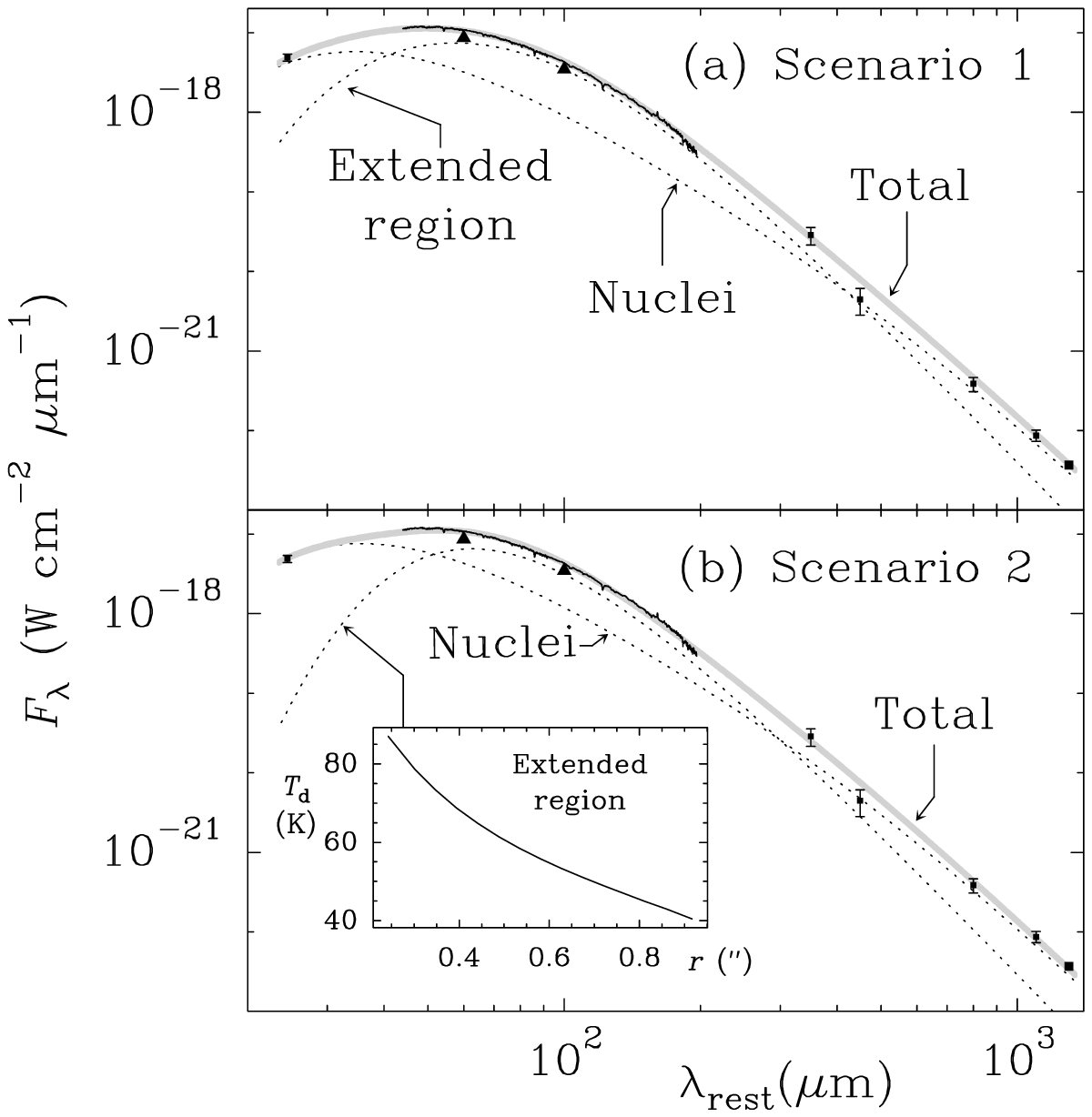}
\caption{Fits of the continuum emission from Arp 220 in the 25-1300 $\mu$m
range for scenarios (a) S$_1$ and (b) S$_2$. 
Solid line shows ISO-LWS spectrum of 
Arp 220, and filled triangles show the 60 and 100 $\mu$m IRAS fluxes for 
comparison. Filled squares show the fluxes measured by \citet[24.5 $\mu$m]{soifer99}; 
\citet*[450 $\mu$m]{eales89}; \citet*[350, 800, and 1100 $\mu$m]{rigo96}; and 
\citet[1300 $\mu$m]{saka99}. Dotted lines 
indicate the computed contributions from the nuclei and the ER, whereas the 
solid grey line show the expected total flux. The insert panel shows the 
calculated dust temperatures in function of the radial angular distance for 
S$_2$ (see text).
\label{fig:cont}}
\end{figure}

Our best fit to the continuum using model S$_1$ assumptions is presented
in Fig. \ref{fig:cont}a, with derived physical parameters 
listed in Table \ref{tab:cont}. In all models, uniform densities throughout the
different components are assumed for simplicity. In Table \ref{tab:cont}, $\lambda_t$
is the wavelength for which the nucleus becomes optically thin ($\tau=1$);
owing to the high opacities involved, the inferred emission is rather
insensitive to the spectral index $\beta$. Thus we have given values of the
physical parameters for $\beta=1.5$ and $\beta=2$. The dust mass has been
derived by assuming a mass-opacity coefficient of 12 cm$^2$/gr at 200 $\mu$m
\citep{hil83}. The parameters that have been allowed to vary in our models of the
ER are the dust temperature $T_d$, the diameter $d$ (within the range
500--800 pc), and the dust opacity; $\beta$ is fixed to 2 to ensure negligible 
emission at millimeter wavelengths (see S99).

{\it Model S$_2$:} In the second scenario (hereafter S$_2$), the emission from the
nuclei at 25 $\mu$m is assumed to be attenuated by foreground dust with
$\tau_{{\rm abs}}(24.5\, \mu{\rm m})=1.2$, a value which was chosen
to be compatible with the silicate absorption observed in S99. The nuclei account
for the required flux at 24.5 and 1300 $\mu$m with $T_d\approx106$ K,
significantly warmer than the $\sim85$ K temperature in S$_1$ and, as before,
the emission in the submillimeter is optically thick (see Table \ref{tab:cont}).
In this scenario as well, however, the flux at 60--100 $\mu$m is again 
underestimated, and an emitting ER must also be involved to account for it. 
Nevertheless, the luminosity from the warm nuclei in S$_2$ is enough to account 
for the observed total luminosity from Arp 220, so that the ER merely re-radiates 
the emission from the nuclei and no extended starburst is then needed
to provide the bulk of the luminosity.

Figure \ref{fig:cont}b shows our best fit for S$_2$. The unique parameters
that have been allowed to vary in our models of the ER are the diameter $d$ and
the dust opacity; $\beta$ is again fixed to 2, and the dust temperature
$T_d$ (shown in the insert panel of Figure \ref{fig:cont}b) has been
computed from the requirement that the heating balance the cooling
throughout the source. The calculation of $T_d$ is carried out by
assuming spherical symmetry, with the nucleus, the primary heating source,
located at the center of the ER. The ER is
divided into a set of spherical shells to account for the variation of $T_d$
with the distance to the nucleus. Once the $T_d$ profile is calculated,
the flux contributions of the attenuated nucleus and the ER are computed
separately, and added up to give the total flux. Despite the good fit to the 
continuum in Fig. \ref{fig:cont}b, S$_2$ implicitly supposes a lack of spherical
symmetry (e.g. the nuclear disk by SYB97) or some clumpiness, because
the derived radial opacity of the ER at 24.5 $\mu$m is 11.3, whereas the
adopted opacity of the absorbing shell in front of the nuclei is
$\tau_{{\rm abs}}(24.5\, \mu{\rm m})=1.2$.
Furthermore, we do not rule out the possibility that the ER is only
partially responsible for the foreground dust absorption of the nuclear
emission by foreground dust.
If the ER were concentrated in a thin disk as proposed by SYB97, little dust
in the ER would be expected to lie in front of the nuclei and significant
dust absorption would be attributed to another component,
``the halo'' (see section \ref{sec:sgrb2}). Therefore our results for 
S$_2$, which assume that the nucleus and the ER are spherically symmetric, 
should be considered only approximate, but suggestive. The intrinsic geometry that 
underlies S$_2$ departs from spherical symmetry and implies that the 
total luminosity, which coincides with the 
luminosity of the nuclei, is lower than the value inferred in S$_1$, where 
spherical symmetry and uniformity is 
assumed for each component (Table \ref{tab:cont}).

Table \ref{tab:cont} shows that $\beta=2$ yields a dust mass of $\sim10^8$ 
\Msun\ for the nucleus, and that S$_1$ gives a mass 1.4 times higher than S$_2$. 
Since SYB97 infer a dynamical mass of $6-8\times10^9$ \Msun\ enclosed in 
the inner 250 pc radius, and this region contains the nuclei and most of 
the inner disk (the ER), S$_2$ with $\beta=1.5$ is favoured in our models 
provided that the gas-to-dust mass ratio is not lower than the standard 
value of $\approx$100. The consistency between the dynamical mass and
the mass derived from the dust emission indicates that the ISM dominates the
dynamics in these inner regions of Arp 220.

It is worth noting that these models may be applied to the source as a whole,
as implicitly assumed above, or alternatively to each one of an ensemble of
$N_c$ smaller clouds of radius $R_c$ that do not spatially overlap along the
line of sight. The value of $N_c\times R_c^2$ determines the absolute
scale, and the radial opacity and temperature distribution of each cloud
as a function of the normalized radial coordinate, $R_c$, determine the
continuum shape. Identical results are found as long as the above parameters
remain constant.
Furthermore, both alternatives give identical total masses, but differ
in the inferred mean density, which scales as $\sqrt{N_c}$.
$N_c=1$ gives the lowest mean density $<n({\rm H}_2>$, which
is listed in Table \ref{tab:cont}. Typical values of a
few$\times10^4$ \cmt\ are derived for the nuclei, accounting for the emission from
molecules with high dipole moment such as HCN, HCO$^+$, HC$_3$N and CN (see also
SYB97). For the ER we obtain $n({\rm H}_2)<10^3$ \cmt; since HCN and HCO$^+$ appear to
show extended low-level emission \citep{rad91b}, it is suggested that the actual
density is higher than this lower limit or that the gas is clumpy.
The low density derived for the ER may also be a consequence of the
spherical shape attributed to the ER: if the mass we derive for the
ER were concentrated in a thin disk with full thickness of 32 pc (SYB97),
the mean H$_2$ density would be $7\times10^3$ \cmt. We will adopt the
density given in Table~\ref{tab:cont}, $n({\rm H_2})=5.3\times10^2$ \cmt, in the 
models for molecules and atoms, but will also explore the results obtained with a 
density one order of magnitude higher than the quoted value.

Although both scenarios S$_1$ and S$_2$ reproduce the continuum 
emission from Arp 220 over the 25-1300 $\mu$m interval and support 
the constraints on the nuclear sizes derived from the available high angular 
resolution continuum measurements, the dynamical masses inferred 
from CO millimeter line observations favour S$_2$ with $\beta=1.5$ for the nuclei. 
Moreover, as we discuss in sections \ref{sec:oh} and \ref{sec:discussion:er},
the observed line absorption/emission also favours model S$_2$. We thus adopt
scenario S$_2$ for the detailed modeling and analysis of the line emission and 
absorption.

\subsection{Extinction}
\label{sec:extinction}

In both scenarios, the high brightness and compactness of the nuclei imply extreme 
continuum optical depth, corresponding to $A_V\sim10^4$ mag. This conclusion 
is in strong contrast with the much more moderate values derived from infrared 
and radio hydrogen 
recombination lines \citep{genzel98,anant00}. The high extinction derived here is 
the direct result of the measured 1.3 mm continuum flux from the nuclei, 210 mJy
\citep{carico92,saka99}, and the observed upper limit of the corresponding
source size, $\sim0\arcsdot4$ \citep{saka99}. These values imply
$\tau_d^{1.3{\rm mm}}\times T_d({\rm K})\sim40$, which shows that even
assuming unexpectedly high average dust temperatures (e.g. $T_d=200$ K) and
$\beta=1$ the dust emission is still optically thick even at 200 $\mu$m.
On the other hand, the radio recombination lines observed by \citet{anant00}
are not affected by dust obscuration, although their predicted fluxes and the derived
extinction may be somewhat model dependent.
These very different extinction values may be understood if we assume that
the observed H recombination lines, tracing primarily star formation, are
formed in the outermost regions of the nuclei, while a buried central
energy source, responsible for the heating of dust in the innermost regions 
of the nuclei and a significant fraction of the galactic luminosity, is weak 
in recombination lines.
If weak in recombination lines, the buried energy source is presumably weak
in PAH features and PDR lines as well, consistent with the strong [C II] deficit 
in Arp 220 (see also section 
\ref{sec:modelmol}). Dust-bounded ionized regions, in which most of the Lyman
continuum from nuclear starbursts or AGN is absorbed by dust rather than by gas, 
may explain these properties of Arp 220, as was proposed by \citet{luh03}.
We further argue that the Lyman continuum luminosities derived from recombination 
lines do not empirically rule out the possibility that an AGN accounts for 
more than $\sim50$\% of the bolometric luminosity of Arp 220, because of the high 
dispersion of $L_{{\rm bol}}/L_{{\rm Lyc}}$ values shown by both 
starburst galaxies and AGN, the range of $L_{{\rm Lyc}}$ values derived from 
different tracers in Arp 220, and the uncertainties in the assumed
extinction law and the derived extinction \citep{genzel98}. Our derived 
$N$(H$_2$)$\sim10^{25}$ 
\cmd\ is high enough to obscure a source of high 5--10 keV luminosity from
one or both nuclei in Arp 220, so that a hidden AGN is allowed despite 
the relatively weak X-ray luminosity observed in Arp 220 \citep{clem02}.
\citet{haas01} have also argued that a hidden AGN powers much of the 
luminosity of Arp 220 on the basis of the observed submillimeter continuum 
excess relative to the 7.7 $\mu$m PAH flux.

\section{Models for molecules and atoms} \label{sec:modelmol}

\subsection{Comparison with Sgr B2} \label{sec:sgrb2}

\begin{figure*}
\epsscale{1}
\plotone{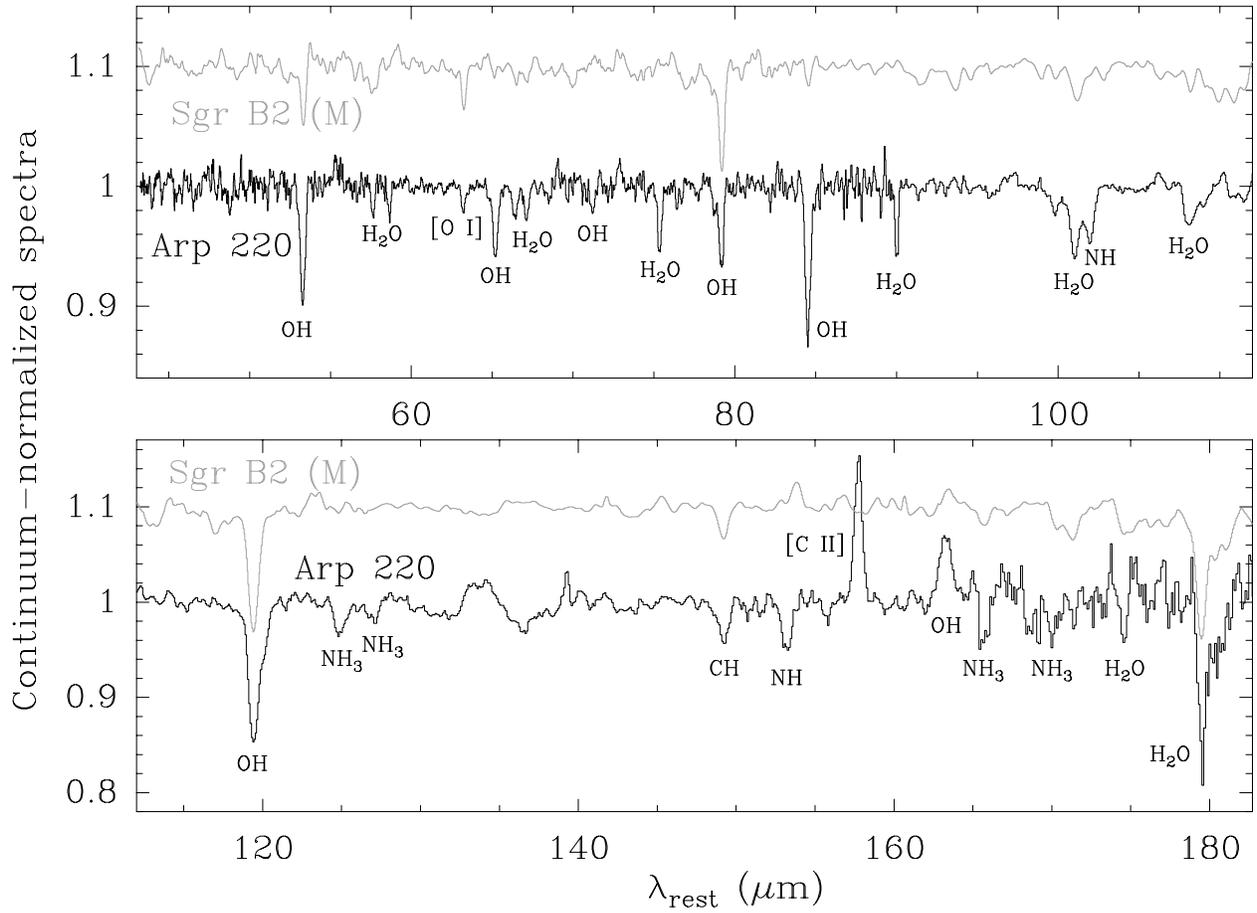}
\caption{Continuum-normalized spectra of Sgr B2 (M) and Arp 220. The main
carriers of some line features are indicated.
\label{fig:arpsgr}}
\end{figure*}

The comparison of the spectrum of Arp 220 with that of some well-studied
galactic sources provides important clues about the regions where the
observed lines are formed, while emphasizing the unique features that
characterize the extragalactic source. In this sense, Sgr B2 (component
M) is an ideal comparison source, because it shares common observational 
properties with Arp 220 despite the obvious differences in spatial scale (and
indeed possibly in nature). Figure \ref{fig:arpsgr} shows the continuum-normalized
spectra of Sgr B2 (M) (kindly provided by J.R. Goicoechea) and Arp 220. Sgr B2
harbors newly born OB stars, ultracompact H II regions, and hot cores, enshrined
in a dusty envelope which is heated by shocks and by radiation,
and which radiates a high IR luminosity (see GRC04 and references therein).
The envelope has the highest extinction in the direction of the N- and
M-condensations, with optically thick emission in the far-infrared up to $\sim$200
$\mu$m, and with foreground absorption lines of OH, \hdo, CH, and [O I] observed
with the ISO/LWS grating (GRC04, Fig. \ref{fig:arpsgr}).
Fabry-Perot observations of Sgr B2 have also allowed the detection of other
molecular species like NH, \nht, \nhd, HD, H$_3$O$^+$, and C$_3$, as well 
as high excitation lines of OH and \hdo\ (GRC04 and references therein). As shown
above (section \ref{sec:cont}), the extinction toward the nuclei of Arp 220
is also very high, although the dust is significantly warmer than in the 
Sgr B2 envelope. The comparison between both sources is meaningful 
(at least as a first approximation, and from the point of view of the
radiative transfer) as long as the nuclear region of Arp 220 can be considered 
an ensemble of continuum-thick molecular-rich clouds such as Sgr B2. For
an ensemble of Sgr B2-like clouds, since both the continuum and the 
line absorption scale with the number of clouds, the
{\em continuum-normalized} spectrum is the same as that of
one individual cloud. This result applies even if the lines in the ensemble
are broadened relative to the one-cloud emission due to cloud-to-cloud
velocity dispersion and rotation, provided that the lines remain unresolved with 
the grating resolution. Thus the differences between the two spectra in Fig. 
\ref{fig:arpsgr} reveal real differences in excitation and/or column 
densities.

The high-excitation lines of OH and \hdo\ are much stronger in Arp 220 than
in Sgr B2 (M) (see Fig. \ref{fig:arpsgr}). In particular, the OH
$\Pi_{3/2}\, J=9/2-7/2$ 65 $\mu$m line, with strong absorption in Arp 220,
is not detected in the grating spectrum of Sgr B2, and the OH
$\Pi_{3/2}\, J=7/2-5/2$ 84 $\mu$m line is also much weaker in Sgr B2 (M). This
strongly indicates the presence in Arp 220 of a high excitation region with 
relatively high OH column densities. The high spectral resolution 
($\Delta v\sim$35 \kms) of LWS Fabry-Perot mode observations allowed 
\citet{goi02} to detect high excitation OH lines in Sgr B2, and showed that 
they are pumped through absorption of far-infrared photons. 
In Arp 220 also, these lines appear to be pumped by the strong infrared
radiation flux in the neighbourhood of the nuclei 
(section \ref{sec:outline}).
Toward Sgr B2 (M), \citet{goi02} derived $N$(OH)$\approx2\times10^{16}$ \cmd,
and we may expect significantly higher column densities toward the nuclei of
Arp 220. The peculiarity of Arp 220 is also revealed by the relatively strong
absorptions in the NH$_3$ and NH lines. These species have also been
detected toward Sgr B2 (M) with Fabry-Perot spectral resolution 
\citep[GRC04]{cecca02}, but Fig. \ref{fig:arpsgr} indicates much higher column 
densities in Arp 220, at least toward the nuclei.

GRC04 found that the OH $\Pi_{1/2}\, J=3/2-1/2$ 163 $\mu$m line, pumped through
absorption of photons in the OH $\Pi_{1/2}-\Pi_{3/2}\,J=3/2-3/2$ 53.3 $\mu$m
line, shows emission over a large region associated with Sgr B2. In Arp 220 the
line is strong, suggesting significant widespread emission, i.e. from the
ER. It is also worth noting that, although the [C II] line is not
detected in the grating spectrum of Sgr B2 (M), Fabry-Perot observations
allowed its detection \citep[GRC04]{vastel02}, with a flux of
$\approx1.5\times10^{-17}$ W \cmd\ for the component observed in emission. 
This value is, within a factor of 2, similar to strengths in the surrounding region
where the continuum is however much weaker (GRC04).  Therefore, in addition to 
effects of self-absorption and absorption of the continuum by C$^+$ in foreground
excitation clouds \citep[GRC04]{vastel02}, the low [C II]/FIR ratio at Sgr B2 (M)
is due to a strong increase in the FIR emission
that is not accompanied by a corresponding rise of PDR line emission.
Dust-bounded ionized regions \citep{luh03}, together with extinction effects in the
far-infrared, might account for the lack of PDR line emission associated with this
additional infrared component. This is similar to our hypothesis for
Arp 220's nuclei (section \ref{sec:cii}). The [C II] 
emission in Arp 220 is expected to arise from PDRs in the ER, where the bulk of 
the PAH emission is found \citep{soifer02}. As shown in Fig. \ref{fig:cont}, the 
ER dominates the observed far-infrared emission, but since the main heating sources 
are, according to model S$_2$, the nuclei, the ER is mainly heated via absorption 
of infrared rather than UV photons, its {\em intrinsic} luminosity is relatively 
low, and hence the [C II] line remains weak (section \ref{sec:cii}).

Finally, we stress that Arp 220 and Sgr B2 (M) show similar strong
absorptions in the ground-state lines of OH and \hdo, as well as in the CH line 
at 149 $\mu$m and in the [O I] at 63 $\mu$m. Fabry-Perot observations 
of these lines toward Sgr B2 (M) 
\citep[GRC04]{cerni1997,stacey87,bal97,vastel02},
indicate that most of the absorption is produced at velocities from $-100$
to 30 \kms\ associated with diffuse low-excitation clouds located along the 
line of sight to Sgr B2 but not physically associated with it. 
The similar absorptions found in Arp 220 strongly suggest 
the presence of such a diffuse, absorbing medium. 
In fact we have found that the combination model of the nuclei and the
ER that reasonably reproduces the high excitation OH and \hdo\ lines, fails
to explain the strong absorptions in the lowest-lying OH and \hdo\ lines.
We suggest that in Arp 220, this diffuse, absorbing component is also
responsible for the low ratio of the CO (2-1) line to 1.3mm continuum emission 
toward the nuclei of Arp 220 \citep{saka99}, and for the low brightness of 
both the CO (1-0) and CO (2-1) lines (SYB97), which indicate line-of-sight 
blocking of the nuclear CO emission by low-excitation gas \citep{saka99}.
The above points strongly suggest the presence of an 
absorbing component in front of the nuclear region in which {\em both} the 
particle density and the infrared radiation density are relatively low. This 
component could also account for significant absorption of the nuclear 
continuum emission in scenario S$_2$ if the ER were a thin disk 
(section~\ref{sec:cont}). We will refer to this diffuse component of Arp 220 
as ``the halo''.

\subsection{Outline of the models} \label{sec:outline}

The dust models described in section \ref{sec:cont} set up the basis for the molecular 
calculations. These are carried out with the method described in 
\citet{gon97,gon99}, 
which computes the statistical equilibrium populations of a given molecule by assuming 
spherical symmetry and line broadening caused by microturbulence and/or radial velocity 
gradients. In the present calculations we have assumed, for simplicity, pure 
microturbulent line broadening, but some tests showed that the inclusion of
a radial velocity gradient hardly modified the results. Rotational motion, 
which is present around the nuclear region of Arp 220 
\citep[e.g. SYB97,][]{saka99,ds98}, is not included. Nevertheless, 
the assumption of spherical symmetry and our neglect of steep
velocity gradients that may result from cloud-to-cloud velocity dispersion
may be considered more critical.

Our non-local code accounts for radiative trapping in the molecular lines,
collisional excitation, and excitation through absorption of photons emitted
by dust. The dust parameters derived above for S$_1$ and S$_2$ are used in the
calculations for molecules. As in the case of the continuum models, we have 
modeled the nucleus and the ER separately. This is required, in S$_2$, by the 
relatively low dust opacities of the ER toward the nucleus as compared with 
the radial opacity of the ER (see section~\ref{sec:cont}), and involves 
inevitably an additional uncertainty.

Since the dust in the nucleus is very optically thick throughout the ISO
wavelength range, model results are only sensitive to the molecular column
densities in the outer parts of the nucleus. It is therefore assumed
that only in the outer regions of the nucleus, where the infrared lines
are formed, the molecular abundances are different from zero 
\citep[see][for the case of Sgr B2]{goi02}. Since dust and molecules are assumed 
to be coexistent, extinction effects within the nucleus are implicitly taken into 
account; they place important constraints on the molecular abundances. We have 
adopted a molecular shell thickness of $2\times10^{18}$ cm, which for a mean
\nhdos$=4.6\times10^4$ \cmt\ (in S$_2$) corresponds to $A_V\sim50$ mag and
$\tau\,(50\,\mu{\rm m})\approx0.3$. For lines around 50 $\mu$m, the contribution 
to the absorption by molecules located deeper into the nucleus was checked in 
some tests to be relatively weak, due to both dust and molecular line optical 
depth effects. Foreground extinction in S$_2$ was also taken into account.
In the models for the ERs, which have much lower continuum opacities, we have
assumed that dust and molecules have uniform abundance ratio throughout the 
whole region. The presence of the central nucleus is included in the calculation 
of the statistical equilibrium populations, but ignored in the calculation of the 
emergent fluxes to avoid accounting for it twice.

Toward the nucleus, absorption of continuum radiation determines the excitation
of OH and \hdo; the radiative rates are much higher than the collisional ones
even for very high pressure molecular gas, such as is found in
molecular shocks (\nhdos$=5\times10^6$ \cmt\ and \tk$=300$ K; 
hereafter we refer to these values as ``shock conditions'' for brevity). 
The rate coefficients of \citet*{offer94} and \citet*{green93} were used
to check the collisional excitation of OH and \hdo, respectively.
If widespread shock conditions were present, only the absorption of the
lowest-lying lines would be significantly affected. Since we use a 
halo to match these lines, we cannot distinguish between shock and
non-shock conditions (i.e., the line ratios are not sensitive to \nhdos\ and 
\tk\ within plausible values). For simplicity, physical parameters for 
the halo are derived by assuming non-shock conditions for the nucleus.
Concerning the ER, widespread shock conditions are not applicable because they 
would involve strong emission in the \hdo\ $2_{12}-1_{01}$ and $3_{03}-2_{12}$ 
lines and in most OH lines (hard to cancel by any halo).  
Our molecular data are therefore only sensitive to the radial molecular column 
density $N$ and the microturbulent velocity dispersion $\sigma_v$, so that only 
these two computational parameters are required to define a model of a given 
component.

Similar to the dust models, models for
molecules may be applied to the source as a whole, or alternatively to each one
of an ensemble of smaller clouds that do not spatially overlap along the
line of sight. Besides the scaling relationships pointed out above, the
molecular column density must remain the same to obtain identical results
when varying $N_c$. On the other hand, once the number of continuum sources
$N_c$ is fixed (e.g., $N_c=1$), the same line fluxes are obtained if
both $N$ and $\sigma_v$ are divided by the same factor $f_c$, and the
resulting fluxes are then multiplied by $f_c$. The latter reflects the 
approximate equivalence between one absorbing cloud with column density $N$ 
and velocity dispersion $\sigma_v$, and $f_c$ clouds, with parameters $N/f_c$ 
and $\sigma_v/f_c$, which overlap on the sky but not in the line-of-sight 
velocity space\footnote{In fact the models could be defined in terms of the 
two independent variables $N/\sigma_v$ and $N\times f_c$; nevertheless we will  
use the variables $N$ and $\sigma_v$ with $f_c$=1.}.

Variations in $N$ and $\sigma_v$ have different effects on the line absorptions.
If $\sigma_v$ decreases the absorptions are weaker for optically thick lines,
but due to the increase of line opacities, the high-energy levels become more
populated and the above weakness is more pronounced for the low-lying lines. On
the other hand, the increase of $N$ has little effect on very optically thick 
lines (most of them low-lying lines), but a larger effect on those with 
moderate opacities.

We have generated a grid of models for the nucleus and the ER of both S$_1$
and S$_2$, by varying the above free parameters $N$ and $\sigma_v$. In each 
model, the molecular shell is divided into a set of sub-shells in order to 
account for the spatial variation of the excitation temperatures of the lines. 
First we searched for the nucleus+ER combination model that best matches the OH and 
\hdo\ non-ground-state lines, with the same value of 
$\sigma_v$ for both species. As pointed out above, the relatively deep 
absorptions of the OH and \hdo\ ground-state lines could not be fitted 
satisfactorily by any model, and a halo component was added to match these
lines. The halo was assumed to be a purely absorbing shell; although
the equilibrium populations were computed in spherical symmetry assuming a 
size of three times that of the ER, limb emission (i.e. emission for impact
parameters that do not cross the continuum source) was ignored in the
calculation of the fluxes emergent from the halo.

Once the model for OH and \hdo\ is determined, models for the other detected 
species are performed, also keeping fixed the value of $\sigma_v$ derived
above for each component. Figure \ref{fig:arp220model} compares the derived 
model spectrum and the observed one, and Table \ref{tab:arp220model} lists the 
inferred parameters. The next sections are devoted to explaining the details 
of these calculations and results.

\subsection{OH and H$_2$O} \label{sec:oh}

Our best model fits for the high excitation lines of OH and \hdo\ come
from model S$_2$, and the results presented here will focus on this scenario.
The main difference between S$_1$ and S$_2$ consists in the higher column densities 
and/or broader line widths required by S$_1$ to reproduce the lines, 
owing to the fact that $T_d$ is significantly lower in S$_1$.
Models for the nucleus with broad line widths ($\sigma_v>60$ \kms, see below), 
however, predict strong absorptions in some OH and \hdo\ lines, such as the OH 
$\Pi_{1/2}\,\, J=5/2-3/2$ 98 $\mu$m and the \hdo\ $3_{13}-2_{02}$ 138 $\mu$m lines, 
which are not observed. Therefore S$_2$, which still requires high column densities, 
is a better fit to the data.

Our best model fits for the high excitation lines involve column densities
of $2-6\times10^{17}$ \cmd\ for both OH and \hdo, and $\sigma_v=50-30$ \kms,
towards the nucleus. The model for the nucleus reproduces nearly the whole
absorption in the OH $\Pi_{1/2}\,\, 7/2-5/2$ and $\Pi_{1/2}-\Pi_{3/2}\,\,5/2-5/2$
lines, most of the OH $\Pi_{3/2}\,\, 9/2-7/2$, and significant absorption in the
other lines but by far too weak in the ground-state lines. It also 
reproduces the full absorptions in the \hdo\ $4_{22}-3_{13}$, $4_{32}-3_{21}$,
$3_{30}-2_{21}$, $3_{31}-2_{20}$, and $4_{23}-3_{12}$ lines, and significant
absorption in the $3_{21}-2_{12}$ and others. The somewhat low value derived
for $\sigma_v$ is required to keep the absorptions weak in some low excitation 
lines which are marginally or not detected. Since low $\sigma_v$ implies 
low velocity coverage for absorption of the continuum, the
column densities that are needed to explain the absolute values of the
absorptions in high excitation lines are relatively high. We stress that this
value of $\sigma_v$ must be interpreted as a strong lower limit on the linewidths
that would be observed with high enough spectral resolution, because
rotation and cloud-to-cloud velocity dispersion would broaden the observed
lines. The CO kinematic models of Arp 220 by SYB97 found a generic value of
$\sigma_v=90$ \kms; as the authors discuss this value should be considered the
joint effect of the local linewidth and the cloud-to-cloud velocity dispersion
over a scale of $\sim$100 pc. Our $\sigma_v$ is the local linewidth involved in 
the calculation of opacities and directly related to the molecular excitation, and 
thus the kinematic value of SYB97 must be considered here an upper limit. 

It is worth noting that the above column densities are derived by forcing the OH
and \hdo\ high excitation lines to arise in the same region. A slight improvement
to the fit of the \hdo\ lines is obtained with even lower $\sigma_v$, 25 \kms,
and $N$(\hdo)$\sim10^{18}$ \cmd. This may indicate that \hdo\ and 
OH toward the nucleus do not arise in the same regions. We will adopt in 
the following the nucleus model with $\sigma_v=50$ \kms, corresponding to
the spectrum shown in Fig. \ref{fig:arp220model} and the parameters given in Table
\ref{tab:arp220model}, and estimate an uncertainty of a factor of
$\sim$3 on the derived column densities.

Once the nuclear component is fit, a search for the combination nucleus+ER
that best accounts for the remaining flux in non-ground-state lines is
carried out. For densities in the ER, $n({\rm H_2})<10^4$ \cmt\
(section~\ref{sec:cont}) and $T_k=T_d$, the OH and \hdo\ collisional excitation
is found to be negligible in comparison with the radiative excitation. We
have also explored the most plausible situation that the OH lines are formed within 
the C$^+$ region of PDRs (see section \ref{sec:cii}): assuming $T_k=300$ K (the
maximum allowed by the collisional rates of \citealt{offer94}) and
$n({\rm H_2})<10^4$ \cmt, radiative rates still dominate over collisional
rates, and results are found indistinguishable from those
obtained with lower $T_k$ values. Only densities above $10^5$ \cmt\ with
$T_k=300$ K would give results significantly different for the
ground-state transitions of OH.

The ER mainly accounts for the OH $\Pi_{1/2}\,\, 3/2-1/2$ 163 $\mu$m line,
which is uniquely observed in emission (but predicted in absorption 
in the nuclei), for more than half of the absorption observed in the OH
$\Pi_{3/2}\,\, 7/2-5/2$ line, for reemission in the $\Pi_{1/2}\,\, 5/2-3/2$, and 
for significant absorption in the three ground-state OH lines.
We estimate for the ER $N$(OH)$\sim2\times10^{17}$ \cmd, with $\sigma_v=50$ 
\kms\ throughout most of the ER and $\sigma_v=90$ \kms\ just around the
nucleus. This higher value of $\sigma_v$ was required to obtain significant
reemission in the non-detected OH $\Pi_{1/2}\,\, 5/2-3/2$ line; since here
geometrical effects may be important, this result should be considered with
caution. For \hdo\ we obtain a significantly lower $N$(\hdo)$\sim3\times10^{16}$
\cmd, giving significant absorption in the $3_{21}-3_{12}$, $2_{20}-1_{11}$, and
$2_{21}-1_{10}$ lines, and some reemission in the $3_{03}-2_{12}$ line.
Since the halo also yields some absorption in these \hdo\ lines but much deeper 
absorption in the ground-state $2_{12}-1_{01}$ one (see below), the relative 
\hdo\ column density in these components is not well determined.

In the halo, the values of $N$(OH) and $\sigma_v$ were determined by
fitting the missing absorption in the three OH ground-state lines. A value of 
$\sigma_v$ relatively low, 15--20 \kms, and $N$(OH)$\sim2\times10^{16}$ \cmd, 
were found to reasonably fit the fluxes of the 79 and 119 $\mu$m lines, 
though the flux of the 53 $\mu$m line is somewhat underestimated 
(Fig. \ref{fig:arp220model}). 
Higher values of $\sigma_v$ would predict too much absorption in the already
saturated 119 $\mu$m line. For \hdo\ we find $N$(\hdo)$\sim1.5\times10^{16}$ 
\cmd. Owing to the strong radiation field from the nucleus and ER, 
\hdo\ in the halo is still significantly excited, thus yielding also some 
absorption in the $2_{20}-1_{11}$, $2_{21}-1_{10}$ and $3_{03}-2_{12}$ lines.

Despite the generally satisfactory fit obtained for the OH and
\hdo\ lines, the shoulder of the OH 119 $\mu$m line, presumably produced 
by the $^{18}$OH $\Pi_{3/2}\,\, 5/2-3/2$ line, is not reproduced. 
The models assume $^{16}$OH/$^{18}$OH=500; however, the isotopic abundance 
ratio in Arp 220 required to reasonably fit the 120 $\mu$m shoulder is as low 
as $^{16}$OH/$^{18}$OH$\sim$50. This value cannot be ruled out if $^{14}$N is 
efficiently converted into $^{18}$O in nuclear processing of high mass stars, 
and then efficiently ejected to the interstellar medium through stellar winds 
and/or supernovae \citep{henkel93}. Also, there is compelling evidence for 
isotopic ratios of 150--200 in starburst regions of nearby galaxies 
\citep{henkel93}. Due to the low spectral resolution of the spectrum, and 
the possibility that the feature is contaminated by other species, we
do not attempt to place useful constraints on this ratio. Nevertheless, we 
conclude that a very low $^{16}$OH/$^{18}$OH abundance ratio likely applies to 
Arp 220, perhaps indicating an advanced stage starburst 
\citep{henkel93}.

The model of Fig.~\ref{fig:arp220model} predicts an absorbing flux of
$1.8\times10^{-12}$ erg s$^{-1}$ cm$^{-2}$ for the
$\Pi_{1/2}-\Pi_{3/2}\,\, 5/2-3/2$ line at 34.6 $\mu$m, in reasonable agreement
with the observed flux of $2.1\times10^{-12}$ erg s$^{-1}$ cm$^{-2}$ reported
by \citet{skin97}. Concerning the OH-megamaser emission at 18 cm,
our models are not suitable to account for them. 
Proper models for the maser emission require inclusion of the hyperfine
splitting, treatment of far-infrared line overlaps, as well as a study of the
influence of the velocity field \citep[e.g.][]{ran95}, which are not 
considered here. The  OH-megamaser emission may be 
radiatively pumped through absorption of photons in the 34.6 $\mu$m and 
53.3 $\mu$m lines, followed by radiative cascade to lower levels.
Since the lower level of the 34.6 $\mu$m transition is the ground-state 
$\Pi_{3/2}\,\, J=3/2$ level, we have found that about 65\% of the modeled absorption 
is predicted to occur in the foreground halo, rather than in the nuclei. As a 
consequence, the pump efficiency that would be required in our model to explain 
the 1.667 GHz OH-megamaser line via the 34.6 $\mu$m line alone is higher
than the value of $\sim1$\% derived by \citet{skin97}. Nevertheless, given the 
uncertainty in the derived nuclear OH column density, the possibility 
that some of the maser emission arises in the most central regions, and the expected 
additional contribution to the pumping by the 53.3 $\mu$m line \citep{bur90}, 
we conclude that our models are roughly consistent 
with the OH-megamaser excitation scheme discussed by \citet{skin97}.

\subsection{CH}

Since CH is close to the Hund's coupling case (b) limit in its $^2\Pi$ ground
state \citep[e.g.][]{brown83}, we denote its levels through ($N$, $J$), where $N$ is
the case (b) rotational quantum number and $J=N\pm\frac{1}{2}$. We assume that the
(2,3/2)--(1,1/2) CH line we observe at 149.2 $\mu$m arises in the halo,
based on the results obtained toward the Galactic Center (see GRC04).
We derive $N$(CH)$\approx2\times10^{15}$ \cmd\ by fitting the feature.
Unlike the case of Sgr B2, however, the submillimeter emission from the 
nuclear region of Arp 220 is strong enough to populate significantly the 
(1,3/2) level via absorption of photons in the (1,3/2)--(1,1/2) line at
560 $\mu$m, so that our models for the halo predict some contribution by 
CH (2,5/2)--(1,3/2) to the absorption feature at 181 $\mu$m (Fig.
\ref{fig:arp220model}). The latter is uncertain, however, because the feature 
at 181 $\mu$m could be contaminated by H$_3$O$^+$ (GRC04) and/or by stronger
absorption of H$_2^{18}$O, whose abundance relative to H$_2^{16}$O could be
enhanced relative to the assumed value of 1/500. On the other hand,
there is a wing-like feature at 118.5 $\mu$m, observed in both the forward
and reverse scans, which could be caused by the doublet CH (3,7/2)--(2,5/2).
If so, and since the excitation of this line requires a relatively
strong radiation field, there would be CH in the nuclei that would
account for about 1/3 of the absorption at 149.2 $\mu$m, and the CH 
column density in the halo would be 2/3 of the quoted value.

\subsection{NH and NH$_3$} \label{sec:amon}

In contrast with OH and CH, most of the column density of NH and \nht\
we model is contained within the nuclei.
Assuming that the 153.2 $\mu$m absorption feature is caused entirely by NH, we
have obtained $N$(NH)$\sim10^{16}$ \cmd\ toward the nucleus. The model
reproduces the marginal absorption feature at 76.8 $\mu$m, attributable to
the $N_J=4_J-3_{J'}$ lines. For the above column density, the $2_2-1_{1}$ and 
$2_3-1_{2}$ lines at 153.2 $\mu$m become saturated and the associated
feature is not completely reproduced, so that we have added an additional
halo component with $N$(NH)$\approx2\times10^{15}$ \cmd\ (Table 
\ref{tab:arp220model}). Nevertheless, this model still underestimates the 
absorption at 102 $\mu$m, strongly suggesting the contribution from other species 
(section \ref{sec:generalresults}).

\begin{figure*}
\epsscale{1.1}
\plotone{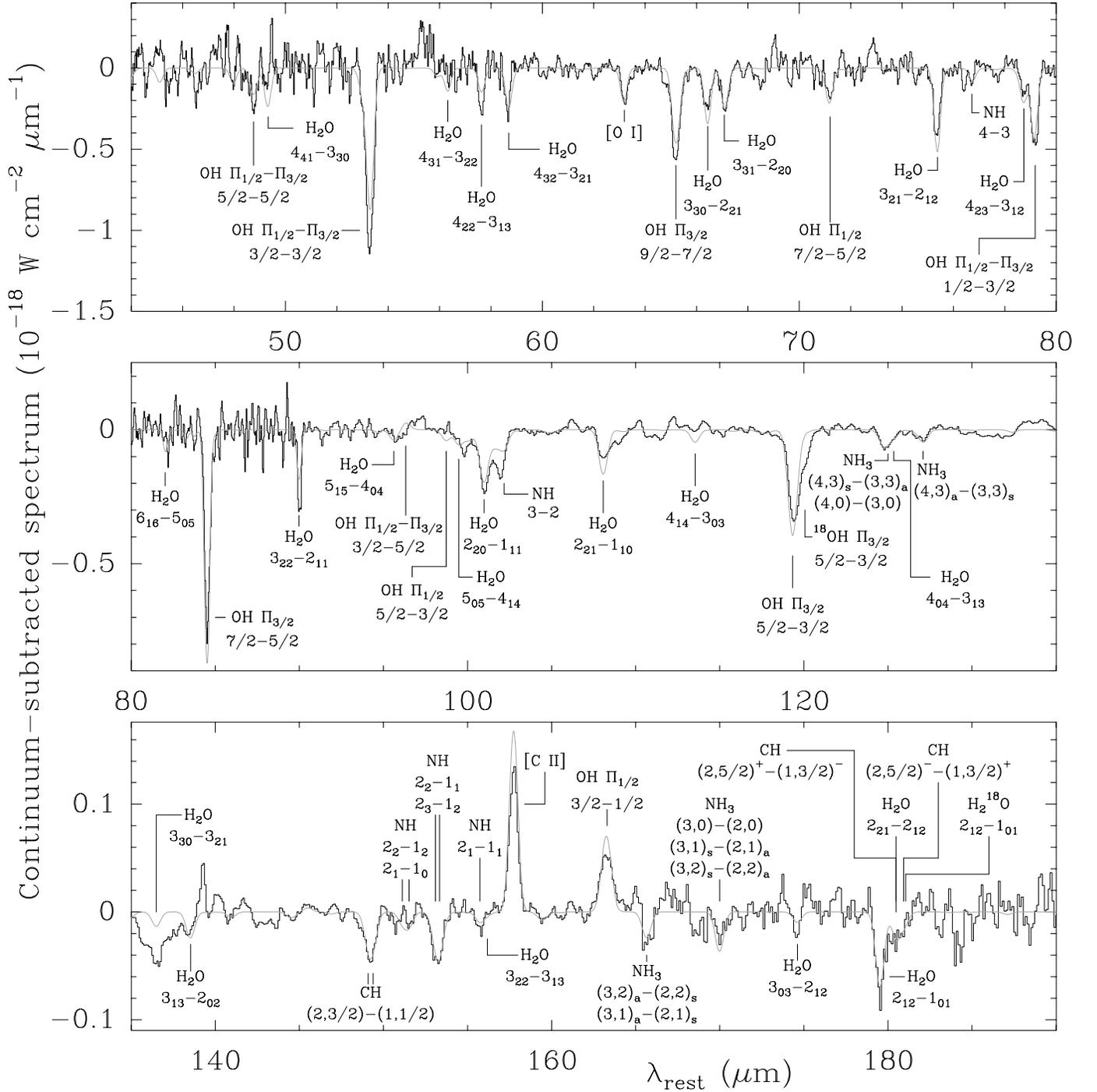}
\caption{Continuum-subtracted spectrum of Arp 220, compared with the result
of the model (in grey). The line features that contribute more to the model
are identified (see text). 
\label{fig:arp220model}}
\end{figure*}

Within a given $K-$ladder, the excitation of the \nht\ non-metastable
levels is determined by absorption of far-infrared continuum photons, while
the metastable levels in $K=2,3,{\ldots}$ are populated through collisions
\citep[see e.g.][for the case of Sgr B2]{cecca02}.
In the model of Fig. \ref{fig:arp220model} we
have assumed an average density of \nhdos$=4.6\times10^4$ \cmt\ and
$T_k=100$ K, but we have checked that the model results are 
insensitive to the adopted $T_k$ because of the blending of lines from
different $K-$ladders to each spectral feature in our spectrum 
(Fig. \ref{fig:arp220}).
We obtain $N$(\nht)$\sim3\times10^{16}$ \cmd\ toward the nucleus 
to fit the absorptions at 125 and 127 $\mu$m (Fig. \ref{fig:arp220model}). 
Besides the absorption in the lines showed in Fig. \ref{fig:nh3lines}, 
the model predicts significant absorption at $\approx$100 $\mu$m and 
$\approx$101.6 $\mu$m, caused by $(J,K)=(5,K)-(4,K)$ lines, which is
still insufficient to account for the observed absorption around 101.6 $\mu$m.
The model also fails to explain the strong absorption at 166 
$\mu$m, and therefore an halo component with $N$(\nht)$\sim4\times10^{15}$ 
\cmd\ has been added to the global model of Fig. \ref{fig:arp220model}.
Nevertheless, the halo components of NH and \nht\ should be considered 
uncertain, because variations in the background continuum associated 
with each component could in principle account for the missing flux in
the lines.

Finally, we have explored the possibility that the \nhd\ radical contributes 
to the spectrum at some wavelengths. The expected strongest absorption 
from \nhd\ is found at $\approx$159.5 $\mu$m, caused by the strongest 
components of the $3_{13}-2_{02}$ ortho line (the hyperfine structure was
neglected in these calculations, but the split of the levels due to the
unpaired electronic spin of 1/2 was taken into account). At this 
wavelength, a marginal absorption feature may be attributed to \nhd, and is
approximately fit with a model for the nucleus where $N$(\nhd)$\sim10^{15}$ 
\cmd. We have included this model of \nhd\ in Fig. \ref{fig:arp220model} to show 
that the expected absorption in other lines, like the $3_{22}-2_{11}$ one at 105 
$\mu$m, do not conflict with the observations, and 
we conclude that $N$(\nhd)$\le2\times10^{15}$ cm$^{-2}$.

\subsection{C II and O I}
\label{sec:cii}

A crucial test of our model is whether it can reproduce the [C II]
157.7 $\mu$m emission and [O I] 63.3 $\mu$m absorption lines.
Among ULIRGs, Arp 220 shows one of the most extreme [C II] deficits 
\citep[$F_{{\rm C II}}/F_{{\rm FIR}}\approx2\times10^{-4}$,][]{luh03}.
The [C II] line is formed within $A_V\le$2 mag from the surfaces of PDRs
\citep*[e.g.][]{wolf90}, where the UV field from nearby high mass stars, or
from the average galactic field has not been significantly attenuated. 
In this region, photodissociation maintains most of the gas in atomic or singly 
ionized form, but some radicals, like OH and NH, find their maximum abundances there 
\citep{stern95}. In particular, OH is expected to be an excellent molecular 
tracer of PDRs' surfaces, given that its abundance is rather low in UV-shielded
quiescent molecular clouds. Its abundance relative to H nuclei within the C$^+$ 
region of dense PDRs is expected to approach the value of $\sim3\times10^{-6}$ 
\citep{stern95}. In fact, \citet{goi02} have found an abundance of
$\approx2\times10^{-6}$ in Sgr B2, and it could be as high as $5\times10^{-6}$
around the galactic center \citep{genzel85}. We have estimated 
$X$(OH)$\sim1-3\times10^{-6}$ toward the nucleus of Arp 220.

On the above grounds, and adopting a gas phase carbon abundance of
$1.4\times10^{-4}$ \citep{ss96}, we assumed $N$(C$^+$)/$N$(OH)=100
in Arp 220 (Table \ref{tab:arp220model}), and we computed the expected 
[C II] line emission by assuming excitation through collisions with 
atomic H \citep{tielens85}. The collisional rates were taken from \citet{lau77a}.
The H densities were assumed to be twice the H$_2$ densities derived 
from the continuum models, i.e. $n({\rm H})=9.2\times10^4$ \cmt\ in the 
nucleus and $n({\rm H})=1.06\times10^3$ \cmt\ in the ER.
Since the critical density is $3\times10^3$ \cmt\ \citep{kauf99},
results for the nucleus are not critically dependent on the assumed density.
They are also insensitive to the assumed temperature
as long as it is higher than $\sim$100 K \citep{wolf90}. Thus our results
are only sensitive to the assumed density in the ER, and to the assumed
C$^+$ abundance.

The result of this calculation has also been added to the overall model of
Fig. \ref{fig:arp220model} ($T_k=500$ K has been assumed). The model
overestimates the observed [C II] emission by only 24\%.
The contribution from the
nucleus is only 13\% of that from the ER because of the low volume
of the nuclear emitting region. The expected line flux from the nucleus could
be lower if absorption of the underlying continuum by low-excitation C$^+$ in
the halo, ignored in this model, occurs as observed in Sgr B2 (M)
(GRC04). The bulk of the line emission arises from the ER. If the
density of the ER were one order of magnitude higher than assumed
(section~\ref{sec:cont}), the modeled [C II] emission would be a factor of
$\approx2$ stronger than in Fig.~\ref{fig:arp220model}. The situation in
Arp 220 resembles what is found in Sgr B2 (GRC04), where the line is emitted 
mainly from an extended region around condensations N and M, 
while the strong FIR source itself is not associated with corresponding observable
[C II] line emission.

The [O I] 63.3 $\mu$m line has been modeled by assuming $N$(O I)/$N$(OH)=250 
(Table \ref{tab:arp220model}); although the oxygen abundance is expected to
be twice that of C$^+$ in the atomic region, it is expected that further
atomic oxygen exists deeper into the clouds \citep{stern95}. 
For this reason, a wide range of excitation temperatures is expected for 
the [O I] line. We have just fitted a single ``effective'' kinetic temperature 
and assumed also collisions with atomic H. The same densities as assumed
above for C$^+$ excitation are used for O I, and the collisional rates are
taken from \citet{lau77b}.

Our calculations show that absorption in the [O I] 63.3 $\mu$m line
is obtained, both toward the nucleus and the ER, with an effective 
$T_k=160$ K, but the line is still too weak to account for the
observed feature. Given the high O I abundances that are expected in diffuse
clouds \citep{bal97}, we have added to the model a halo component with 
$N$(O I)$=3\times10^{18}$ \cmd\ (Table \ref{tab:arp220model}). This  
model accounts for the observed absorption at 63.3 $\mu$m 
(Fig. \ref{fig:arp220model}).

On the basis of the low extinction
derived from infrared and radio H recombination lines and the high 
optical depth derived from our dust models (section \ref{sec:extinction}), 
and the assumption that [C II] emission is expected to suffer extinction 
similar to that of the recombination line emission, we favor the explanation 
proposed by \citet[][see also section \ref{sec:sgrb2}]{luh03} that non-PDR 
far-infrared emission is responsible for the [C II] deficit in Arp 220. 
Nevertheless, our continuum models by themselves (section \ref{sec:cont}) 
cannot rule out the possible role of far-infrared extinction on the measured line 
fluxes: the derived high nuclear far-infrared opacities indicate that
only the outer regions of the nuclei, where the OH and \hdo\ lines 
are formed, and the ER, are able to contribute to the [C II] line emission.
Our model fits cannot discern whether the {\em intrinsic} nuclear 
[C II] emission is negligible or rather is obscured by dust.
In either case, we have shown that the [C II] line is well reproduced 
by assuming that C$^+$ and OH are coexistent.

\section{Discussion} \label{sec:discussion}

\subsection{The Extended Region (``ER'')}
\label{sec:discussion:er}

Our models support the widespread presence of PDRs in the ER. 
Both the high OH and C II column densities indicate that the 
UV field from newly formed stars have a profound effect on the chemistry in 
the ER. Significant contributions from shocks can be neglected,
as was pointed out in section \ref{sec:outline}. The \hdo-to-OH abundance
ratio is significantly lower than 1, probably indicating enhanced \hdo\
photodissociation. As pointed out in section~\ref{sec:cont}, the 
density is uncertain, but thought to be in the range $n({\rm
H_2})=5\times10^2-7\times10^3$ \cmt.

Our models also indicate that the bulk of the [C II] line 
emission arises in the ER. It is therefore likely that star formation is 
responsible for this emission. This result is strongly supported by the
observations of \citet{soifer02}, who found that the PAH emission is also
spatially extended. Assuming a ``normal'' [C II]/FIR ratio, 
i.e. $F_{{\rm [C\,\,II]}}/F_{{\rm FIR}}=5\times10^{-3}$ \citep{stacey91},
the expected {\em intrinsic} FIR emission from the ER is $\sim3\times10^{10}$ 
\Lsun, i.e. $\sim$3\% of the total galactic luminosity. This estimate 
supports the scenario S$_2$ that has been used to model the line emission. 
According to our models, the total FIR luminosity (due to absorption
and re-emission of nuclear infrared radiation) from the ER is much higher
than the intrinsic (PDR) emission. Our estimate of the intrinsic ER
luminosity is somewhat lower than that derived by \citet{soifer02} and 
\citet{spoon04}, who found a lower limit of $\sim7\times10^{10}$ \Lsun, but
confirms qualitatively their results. The starburst luminosity in the ER 
seems to be similar to that of moderately bright infrared galaxies, like NGC 
253 \citep*{radov01}.

The warm mass in the ER derived from our OH models is $\sim10^9$ \Msun, 
assuming $X$(OH)=$10^{-6}$ relative to H nuclei and the warm H$_2$
arising from the same volume as the OH. This estimate of mass may be 
considered an upper limit. From the H$_2$ rotational lines detected by ISO, 
\citet{stu96} estimated $\sim3\times10^9$ \Msun\ of warm gas. Given the uncertain 
correction by extinction, it is likely that an important 
fraction of the H$_2$ emission arises in the ER.

\subsection{The halo}

The models also indicate the presence of a halo. Despite the uncertainties 
in the column densities of this component, they are typical of those 
found toward our galactic nucleus: $N$(\hdo,
OH)$\sim2\times10^{16}$ \cmd\ have been also derived in the diffuse medium toward 
Sgr B2 \citep{cerni1997,neu03,goi02}, and the derived $N$(CH)$\approx2\times10^{15}$ 
\cmd\ is also similar to that found toward Sgr B2 by GRC04 and
\citet{stacey87}. The column densities in the halo derived for NH and \nht\
are uncertain because they are based on single lines, but they could also
exist in a population of molecular clouds located far away from the nuclear region.
Assuming $X$(\hdo)$=10^{-6}$ in the halo \citep{neu00},
$N({\rm H_2})\sim2\times10^{22}$ \cmd\ is obtained and the associated continuum
opacity is $\tau_{200\mu{\rm m}}\sim8\times10^{-3}$. If we further assume a
spectral index $\beta=2$, the derived dust opacity at 25 $\mu$m is $\sim0.5$, 
comparable to the value of 1.2 in scenario S$_2$ (section~\ref{sec:cont}).
Therefore, significant absorption of the nuclear continuum emission is
attributable to the halo.

\subsection{The nuclei}

The similarity of the OH column densities in the ER and the nucleus
may suggest that, at least to some extent, we are observing the same widespread 
OH component, and that OH is more excited toward the nucleus because of the
underlying stronger infrared continuum in that direction. On this ground a PDR
origin of the observed OH would be favoured. However, the inferred high 
\hdo\ column density would be difficult to explain in this context. Although \hdo\ is
expected to form efficiently in UV-shielded regions of dense PDRs
\citep{stern95}, with total column densities similar to those of OH, there
seems to be no clear correlation between $N$(OH) and $N$(\hdo). In fact, the
\hdo\ column density in the ER is significantly lower than that of OH.
Also, detection of \nht\ and, above all, of NH, seem
to indicate an additional nuclear component.

OH could also arise in C-shocks \citep{wat85}, but it is unlikely that they
dominate the OH absorptions because $N$(\hdo) would then be at least one order
of magnitude higher than $N$(OH). However, an interesting possibility is that
those C-shocks, or alternatively hot core regions, are combined with PDRs, 
i.e., \hdo\ produced there is subject to a strong UV field and thus to 
photodissociation. This process may be enhanced if there is hot gas emitting 
X-rays, like in supernova remnants where the OH abundance is expected to be 
$\ge10^{-6}$ \citep{wardle99}, or from a nuclear AGN. In the nuclei of 
Arp 220, several compact 18 cm 
continuum sources indicate the presence of high luminous supernovae 
\citep{smith98}; however, the extended and external OH required to explain the 
infrared data suggests a more widespread component. The diffuse OH megamaser 
emission found in Arp 220 should be related to it \citep{lons98}.
Soft-extended and hard-compact X-ray emission, detected
around and from the nuclei \citep{clem02}, could be responsible for
photodissociation of \hdo\ produced in shocks and hot cores, thus enhancing 
the OH abundance. In particular, hot cores are expected to exist widely in 
the nuclei, given the high dust temperatures and densities found there. 
The presence of J-shocks, where $N$(OH) is expected to be higher than 
$N$(\hdo) except for high enough preshock densities \citep{neu89}, cannot 
be disregarded.

The high column densities obtained for NH and \nht\ seem to indicate that
standard gas-phase PDR chemistry  alone is not able to explain the full molecular 
data in Arp 220. \citet{stern95} predicted for a PDR an NH column density more than 
two orders of magnitude lower than that of OH, whereas we estimate
$N$(OH)/$N$(NH)$\sim$20 in Arp 220. The enhancement of NH relative to OH may be 
more than one order of magnitude. In Sgr B2, GRC04 also found a somewhat high
$N$(NH) relative to OH, i.e. $N$(OH)/$N$(NH)$\sim$30--100. 
It is interesting that the high NH abundance in diffuse clouds is a factor
of $\sim40$ higher than predicted by gas-phase chemical models
\citep{meyer91}. The latter has been used by \citet{wagen93} and
\citet{craw97} to argue for grain-surface production of NH. In principle, this 
process could also help to enhance the NH abundance in Arp 220, because the dust 
in the nuclei has been found to be warm so that grain mantles could efficiently
evaporate, and also because in an enviroment with enhanced cosmic rays, as
expected from a starburst and the consequent high rates of supernovae, or an AGN, the 
release of grain mantles to the gas phase by means of sputtering should be
also enhanced. However, hydrogenation of NH should in principle continue until 
saturation, because indeed we observe \nht\ with a column density twice of NH. 
The issue that now arises is that, if hydrogenation generally completes, then 
a very low NH-to-\nht\ abundance ratio would be expected, and if it does not 
complete like in the models of \citet{wagen93}, the scenario fails because of 
the low relative abundance found for \nhd. In Sgr B2, for example, 
NH:\nhd:\nht=1:10:100  (GRC04), and \nhd\ is found to be fairly abundant 
\citep{vandis93}. 

One possible solution for the low \nhd\ abundance is that the already invoked
enhancement of cosmic rays deeply affects the ion-molecule gas-phase chemistry.
\citet*{feder96} have shown that the high NH abundance found in some diffuse
galactic enviroments could be explained through cosmic ray ionization of atomic
nitrogen, followed by hydrogen abstraction reactions that form NH$^+_n$ and
dissociative recombination that yields NH. The slight endothermicity of 
N$^+$+H$_2$$\rightarrow$NH$^+$+H \citep*{millar97} is not a problem here,
given the high dust temperatures in the nuclei. If in Arp 220 the cosmic-ray
ionization of N is enhanced, the above scheme could give rise to high NH
abundances. Furthermore, both \nhd\ and \nht\ would be much less abundant if
photodissociation is important in those regions. \nht\ would be formed primarily 
in grain mantles through nearly complete hydrogenation, thus again keeping the
\nhd\ abundance low, and released to the gas phase in widespread hot core 
regions relatively shielded from UV fields. \hdo\ could also follow this last 
process.

A chemistry deeply influenced by ion-neutral reactions have been also invoked 
by \citet{aalto02} to explain the high emission from HNC relative to HCN in
Arp 220 and other luminous infrared galaxies. Furthermore, \citet{aalto02}
also found strong subthermal CN emission, indicative of gas at moderate
densities and irradiated by UV fields \citep*{rodfran98}. In the 
simplest scenario, the molecular content of Arp 220 may thus be interpreted 
in terms of hot cores submitted to strong UV and X-ray fields, where enhanced 
evaporation of grain mantles and ion-molecule chemistry induced by cosmic ray 
ionization are also deeply affecting the relative molecular abundances.

\section{Summary} \label{sec:summary}

Due to its high signal-to-noise, the rich far-infrared spectrum of Arp 220 
provides a template for understanding the dusty interstellar medium of
ULIRGs. We have analyzed the ISO/LWS spectrum of Arp 220 using radiative
transfer models applied to both the continuum and line emission. Our main
results are: \\
1. The continuum emission from 25 to 1300 $\mu$m is well reproduced
with a two-component model: (a) the nuclei, with effective size of 
$0\arcsdot4$ and dust temperature of 106 K, which accounts for essentially the whole 
flux at 25 $\mu$m and at millimeter and submillimeter wavelengths, and 
(b) an extended region (ER), whose effective size is $2''$
and which dominates the continuum emission from 60 to 250 $\mu$m. \\
2. The extinction toward the nuclei is very high
($A_V\sim10^4$ mag); the dust in the ER is heated through absorption of
radiation emanating from the nuclei. \\
3. The spectrum of Arp 220 shows molecular lines of OH, \hdo, CH, NH, and NH$_3$,
as well as the atomic [O I] 63 $\mu$m line in absorption and the [C II] 158 
$\mu$m line in emission. The outermost
regions of the nuclei, along with the ER, are traced by the lines observed in
the far-infrared. \\
4. The high excitation lines of OH and \hdo\ are pumped through absorption
of photons emitted by dust. Column densities of $N$(OH)$=2-6\times10^{17}$
\cmd\ and $N$(\hdo)$=2-10\times10^{17}$ \cmd\ are derived toward the nuclei.
In the ER, $N$(OH)$\sim2\times10^{17}$ \cmd\ and $N$(\hdo)$\sim3\times10^{16}$ 
\cmd. We found it necessary to invoke a third component, or halo, to match 
the low-lying lines of OH and \hdo; this halo has column densities that are 
similar to those found toward the Galactic Center 
($N$(OH, \hdo)$\sim1.5\times10^{16}$ \cmd). \\
5. The CH line detected in the far-infrared spectrum of Arp 220 is assumed
to arise from the halo, and the inferred column density is
$N$(CH)$\sim2\times10^{15}$ \cmd. This value is also similar to that found
toward the Galactic Center. \\
6. Models for NH and \nht\ indicate high column densities toward the nuclei,
$N$(NH)$\sim1.5\times10^{16}$ \cmd\ and $N$(\nht)$\sim3\times10^{16}$ \cmd.
The upper limit found for the column density of NH$_2$ is much lower,
$N$(NH$_2$)$\le2\times10^{15}$ \cmd. \\
7. The [C II] 158 $\mu$m line strength is approximately reproduced by assuming
that C$^+$ is 100 times more abundant than OH. Our
models predict that the line arises mainly from the ER, and that 
non-PDR far-infrared emission, with possible extinction effects, 
is mostly responsible for the observed [C II] deficit in
Arp 220. The [O I] 63 $\mu$m line is also matched with an abundance of 250
relative to OH and absorption toward the nuclei, the ER, and the halo. \\
8. PDR molecular chemistry plays a key role in the ER and contributes to
the elevated OH abundance at the expense of \hdo. 
Toward the nuclei, however, important contributions from hot cores, and 
possibly from shocks, is most plausible.  
The nitrogen chemistry, and in particular the high NH 
abundance, seems to be strongly influenced by ion-neutral reactions triggered 
by cosmic ray ionization.



\acknowledgments

We would like to thank the entire LWS and ISAP teams for creating the great
instrument and software that produced the spectrum discussed here.
We are grateful to J.R. Goicoechea for providing the ISO/LWS spectrum of Sgr
B2 and for fruitful discussions, and to L. Colina for useful comments on the
manuscript. E. G-A thanks Spanish SEEU for funding
support under project PR2003-0057, and the Harvard-Smithsonian Center for
Astrophysics for its hospitality. This work has been partially supported 
by NASA Grant NAG5-10659, the NASA LTSA program and the Office 
of Naval Research.

\clearpage


\clearpage


\begin{deluxetable}{cccrrrrrrr}
\rotate
\tablecaption{Physical parameters derived from the continuum emission \label{tab:cont}}
\tablewidth{0pt}
\tablehead{
\colhead{Scenario} & \colhead{Component}   & \colhead{$\tau_{abs}$}   &
\colhead{$d$} & \colhead{$\beta$}  & \colhead{$T_{{\rm d}}$} & 
\colhead{$\lambda_{t}$} & \colhead{$M_d$} & 
\colhead{$<n({\rm H}_2)>$\tablenotemark{a}}  & \colhead{$L$} \\
& & (24.5 $\mu$m) & \colhead{(pc)} & & \colhead{(K)} & \colhead{($\mu$m)} & 
\colhead{(\Msun)} & \colhead{(\cmt)} & \colhead{(\Lsun)} 
}
\startdata
S$_1$ & Nucleus & 0 & 145  & 1.5 & 85 & 640  & $5.1\times10^7$ & 
$6.5\times10^4$ & $4.8\times10^{11}$ \\
 &  &  &   & 2.0 & 85 & 770  & $1.3\times10^8$ & 
$1.6\times10^5$ & $4.8\times10^{11}$ \\
S$_1$ & ER & 0 & 630  & 2.0 & 50 & 110  & $5.2\times10^7$ & 
$8.1\times10^2$ & $8.3\times10^{11}$ \\
S$_2$ & Nucleus & 1.2 & 145  & 1.5 & 106 & 520  & $3.6\times10^7$ & 
$4.6\times10^4$ & $9.8\times10^{11}$ \\
 &  &  &  & 2.0 & 106 & 650  & $9.3\times10^7$ & 
$1.2\times10^5$ & $9.8\times10^{11}$ \\
S$_2$ & ER & 0 & 660  & 2.0 & 40-90\tablenotemark{b} & 82  & $3.9\times10^7$ & 
$5.3\times10^2$ & $-$ \\
 \enddata
\tablenotetext{a}{Assuming a single source instead of an ensemble of clouds}
\tablenotetext{b}{The temperature is calculated throughout the ER from the 
balance between heating and cooling with the nucleus as the primary heating source}
\end{deluxetable}


\begin{deluxetable}{crrr}
\tablecaption{Column densities derived toward Arp 220\tablenotemark{a} 
\label{tab:arp220model}}
\tablewidth{0pt}
\tablehead{
\colhead{Species} & \colhead{Nucleus}   & \colhead{ER}   &
\colhead{Halo} 
}
\startdata
OH & $2.0\times10^{17}$ & $2.0\times10^{17}$ & $1.8\times10^{16}$ \\
\hdo & $2.0\times10^{17}$ & $3.0\times10^{16}$ & $1.5\times10^{16}$ \\
CH & -- & -- & $2.0\times10^{15}$ \\
NH & $1.3\times10^{16}$ & -- & $1.8\times10^{15}$ \\
\nht & $2.8\times10^{16}$ & -- & $4.0\times10^{15}$ \\
C$^+$ & $2.0\times10^{19}$ & $2.0\times10^{19}$ & -- \\
O & $5.0\times10^{19}$ & $5.0\times10^{19}$ & $2.7\times10^{18}$ \\
\enddata
\tablenotetext{a}{Units are \cmd}
\end{deluxetable}

\end{document}